\documentclass[%
 aip,
 amsmath,amssymb,
 reprint,%
]{revtex4-1}

\usepackage[
    type={CC},
    modifier={by},
    version={4.0},
]{doclicense}

\usepackage{graphicx}
\usepackage{dcolumn}
\usepackage{bm}

\usepackage[utf8]{inputenc}
\usepackage[T1]{fontenc}
\usepackage{mathptmx}
\usepackage{amsmath}
\usepackage{graphicx}
\usepackage{amsfonts}
\usepackage{lipsum}
\usepackage{multirow}
\usepackage{bbm}
\usepackage{xcolor}
\usepackage{soul}

\usepackage{relsize}
\usepackage{comment}
\usepackage{amsthm}

\newtheorem{rem}{Remark}
\begin{document}

\preprint{AIP/123-QED}

\title[]{Thermostatic control for demand response using distributed averaging and deep neural networks}

\author{Kshitij Singh}

\email{singh_kshitij@outlook.com}
\author{Pratik Bajaria}%
\email{pkbajaria_p15@ee.vjti.ac.in}
\affiliation{ 
Student/Research Scholar, Veermata Jijabai Technological Institute, Mumbai - 400 019, India
}

\date{\today}

\begin{abstract}
Smart buildings are the need of the day with increasing demand-supply ratios and deficiency to generate considerably. In any modern non-industrial infrastructure, these demands mainly comprise of thermostatically controlled loads (TCLs), which can be manoeuvred. TCL loads like air-conditioner, heater, refrigerator, are ubiquitous, and their operating times can be controlled to achieve desired aggregate power. This power aggregation, in turn, helps achieve load management targets and thereby serve as ancillary service (AS) to the power grid. In this work, a distributed averaging protocol is used to achieve the desired power aggregate set by the utility using steady-state desynchronization. The results are verified using a computer program for a homogeneous and heterogeneous population of TCLs. Further, load following scenario has been implemented using the utility as a reference. Apart from providing a significant AS to the power grid, the proposed idea also helps reduce the amplitude of power system oscillations imparted by the TCLs. Hardware-based results are obtained to verify its implementation feasibility in real-time. Additionally, we extend this idea to data-driven paradigm and provide comparisons therein.
\doclicenseThis
\end{abstract}
\maketitle

%\setlength\linenumbersep{0.12cm}\relax
%\linenumbers
\section{Introduction}\label{sec 1}
Sustainability is the process of maintaining change in a balanced environment, in which the exploitation of resources, the direction of investments, the orientation of technological development, and institutional changes are all in harmony and enhance both current and future potential to meet human needs and aspirations. With the advent of civilization, to serve electrical and transportation needs, fossil fuel reserves have been over-utilized. Excessive use of fossil fuels and increasing human settlements, has brought imbalance to the environment. Thus, researchers have moved towards building settlements that can serve individual requirements, thereby mean no harm to the environment utilizing least natural resources; hence are sustainable. To support the central power grid, ancillary services (ASs) - a bidirectional power flow architecture is employed.
\par ASs support the transmission of electric power from seller to purchaser, given the obligation of control areas and transmitting utilities within the control area, maintaining reliable operation of the interconnected transmission system \cite{hirst1997}. Load following, reactive power voltage regulation, system protection services, loss compensation, system control, load dispatch services, and energy imbalance services are some ASs. Load following is a service program that monitors real-time consumption about scheduled loads to consume what is estimated. In contrast, load dispatch services take real-time action in case of any discrepancies. System protective services are energy reserve that compensates for any discrepancies in the scheduled loads and dispatches where the loss compensation scheme guides its use. The energy imbalance programs take all these actions using system control service. Reactive power voltage regulation refers to services that mitigate perturbation in the voltage online. As observed, these services either fall into demand-side or supply-side management, but with increasing demand to supply ratio, it is advisable to optimize loads to match available resources.
\par ASs have gained attention with increasing penetration of renewable energy sources (RESs) in the course of replacing fossil fuels. Although renewable energy sources (RESs) are abundantly available in nature, they suffer from intermittencies. These intermittencies are significant enough to render a power grid unstable and hence require additional regulating circuits to avoid direct injection \cite{kara2014, yan2017, kara2015, meyn2015}. One of the ways of injection is to store it in a battery and then provide it to the grid using a battery management system (BMS) with a stable discharge rate. Although reliable, this strategy bears a high capital expenditure for installation and hence is commercially less feasible \cite{siano2014, pavic2016, teng2017, malik2018}. Therefore, research has moved in the direction of devising methods to regulate the loads with techno-commercial feasibility. Demand Response (DR) programs have been developed to ally with several handshaking criteria and thereby maintain power grid stability. 
\par DR programs allow consumers to adjust their electricity consumption in response to energy prices or incentive payments. ASs form a major part of these DR programs, and optimal use of these services can come to an advantage in terms of adjusting loads as per utility demands. As per a survey in U.S. \cite{eia2012}, cooling and heating loads account for the largest annual residential electricity consumption. And the scenario remains the same globally, thereby requiring optimization of these loads to attain required power aggregation. Fundamentally, all heating/cooling loads (or TCLs) comprise a temperature transducer, a thermostat, and other assemblies like compressor, and condenser. It can be noted, though, that TCLs follow a scheme whereby they switch between ON and OFF states to attain required setpoint temperatures. ON states consume full capacity power of the TCL, and OFF states imply dormancy. Thus, adjusting these operation states can help achieve the required power aggregate and thus provide load-following ancillary service. 
\par In literature \cite{venkat2016, venkat20162}, field experiments are conducted using domestic household refrigerators to quantify the flexibility of TCLs by advancing or delaying operation times, thereby achieve required aggregate power. To avoid synchronization, parameter randomization of TCLs by compartmentalization of compressor cycles is also an effective scheme \cite{malik2018}. An automatic generation control signal (AGC - a demand-supply balancing technique) tracking \cite{ma2017} and a two-level scheduling protocol \cite{zhou2017} achieves flexible TCL operating cycles in an intra-day electricity market making it a valuable asset to DR programs. However, with increasing heterogeneity in system parameters, the performance worsens in all the schemes mentioned above. On the other hand, a multi-objective model predictive control (MPC) can also be employed for residential heating (or using heat as an energy carrier) with heat pumps \cite{baeten2017, hu2017}. Again, the number of parameters used makes it difficult to be implemented for a large population of TCLs.

\par Aggregated models for TCLs can carry information about the overall power consumption of TCLs and can be framed using discrete state-space dynamics. For instance, a centralized control scheme on an aggregated model can provide effective frequency regulation \cite{hu2017,wu2018}. A Markovian probabilistic framework \cite{zhao2018} or a load shedding strategy \cite{luo2017} can also be used to achieve similar objectives. Concepts from optimization theory related to the tracking device, convex polytopes, and binary multi-swarm, particle optimization, are also used to control the desynchronization of TCLs \cite{iacovella2017,zhao2017,ma2018}. Batch reinforcement learning based on Monte Carlo methods is used to allow TCLs to participate in day-ahead markets, thereby achieve DR objectives \cite{ruelens2017}. A nano-grid solution to store thermal energy in TCLs for later use has also been experimented \cite{burmester2017}. This approach avoids wastage by regulating maximum power point tracking (MPPT) of a photovoltaic panel. In literature \cite{ghanavati2018}, a partial differential equation (PDE) based Fokker-Plank model for demand-side management is also proposed, and as known, requires highly sophisticated tools for solving. Additionally, an event-triggered control of TCLs on a hierarchical architecture is implemented over a network for optimization of communication cost \cite{meng2017}. All of the methods mentioned above are probabilistic, discretized, and involve considerable computation cost.
\par Complexity of population of TCLs increases with the number of TCLs, and with the reference aggregate power set by the utility. The undesired synchronization of TCL duty cycles by the control policies limits the time scale and capacity of ASs that a population can provide. By sequentially changing the set point temperature, safe protocols have been developed to minimize the unwanted power oscillations \cite{sinitsyn2013,mehta2014}. Other ways of providing power regulation include variable deadband strategy \cite{sinitsyn2013}, injecting a delay of $M$-minutes \cite{sinitsyn2013}, and deploying state stacking technique based on priority measures \cite{hao2015,vrettos2012}. These measures are defined as the distance from the switching boundaries to reduction in aggregate power. TCLs in the ON stack closer to switching boundary turn ON first, for an increase in aggregate power; TCLs in the OFF stack closer to the switching boundary are turned OFF first and so on. On similar lines, a phase response curve (PRC) based technique effectively helps to manoeuvre an ensemble of TCLs by using a control input to modulate duty cycles as well as induce delay/advance and thereby changing the set point temperature \cite{bomela2018}. The switching dynamics of TCLs are similar to oscillators, and the phase-dependent model has been realized to achieve the desired objective. A continuum model for studying the synchronization dynamics of TCLs to understand and mitigate potential risks of a decentralized response provider offers deeper insights to TCL dynamics \cite{webborn2018}. All these methods are effective but involve high computation costs and challenges for practical implementation.
\par This study extends the Kuramoto phase oscillator model \cite{bajaria2019dynamic} used to control the phase delay/advance of the operating cycles of the TCLs which helps to achieve permissible load requirements set by the utility. We provide an extensive study on the possibilities of practical implementation of the model discussed in the previous work \cite{bajaria2019dynamic}. Based on our findings, we propose a distributed averaging protocol to achieve similar (and in a better way) objectives obtained using Kuramoto model \cite{bajaria2019dynamic}. Hardware-based outcomes, in conjunction with computer simulations, provide validity to the proposed idea. 
\par Additionally, we extend the idea of phase desynchronization using machine learning (ML) fundamentals to control overall aggregate power consumption of TCLs. The major concern being reduction in computational cost of deterministic approaches to data-driven ones. Some of the articles from literature justify using ML for control of TCLs \cite{kara2014data,kara2014quantifying,costanzo2016experimental,ruelens2019direct,almalaq2017review,kazmi2019multi}. In particular, our views are concurrent with inclusions of deep neural network (DNN) to learn the right phase thereby achieve required power aggregate \cite{ruelens2019direct}. We exploit the characteristics of delay calculator (explained later in the manuscript) and extend it to DNNs to reduce computations every time the network architecture/parameters gets perturbed. 
\par The paper is organized as follows. In section \ref{sec 2}, we lay down fundamentals to develop a platform to propose the mentioned idea. The motivation behind the proposed paper has been discussed briefly in section \ref{sec3}. A novel model for the control of TCLs namely a distributed averaging protocol, is discussed in section \ref{sec 4}. In section \ref{sec 5}, we provide software and hardware outcomes using LEDs as acting TCLs, which simulate similar characteristics, thereby validating the proposed idea. Section \ref{sec 6} presents a competitive benchmark, where we compare the proposed model with those available in the literature and provide its competencies using peer-to-peer reviews. In the last section, we provide a deeper insight to the characteristics of delay calculator and resolves its ``curse of dimensionality" problem using a DNN.

\section{Prerequisites}\label{sec 2}
\subsection{Behaviour of TCLs}\label{subsec1}

\par A TCL switches ON until it reaches the upper limit of the set threshold temperature, post which it turns OFF until reaching the lower limit of the threshold. Usually, TCLs employ a thermostat that measures the surrounding temperature and thereby generates switching signals to regulate the TCL. Typically, a hybrid model is employed to express the dynamics of TCL mathematically. Incorporating the thermal resistance, capacitance, energy transfer rate, and switching signal, the hybrid model describes the temperature dynamics. It can be stated as follows,

\begin{figure}[t!]
\centering
\begin{tabular}{c}
\includegraphics[scale=0.15]{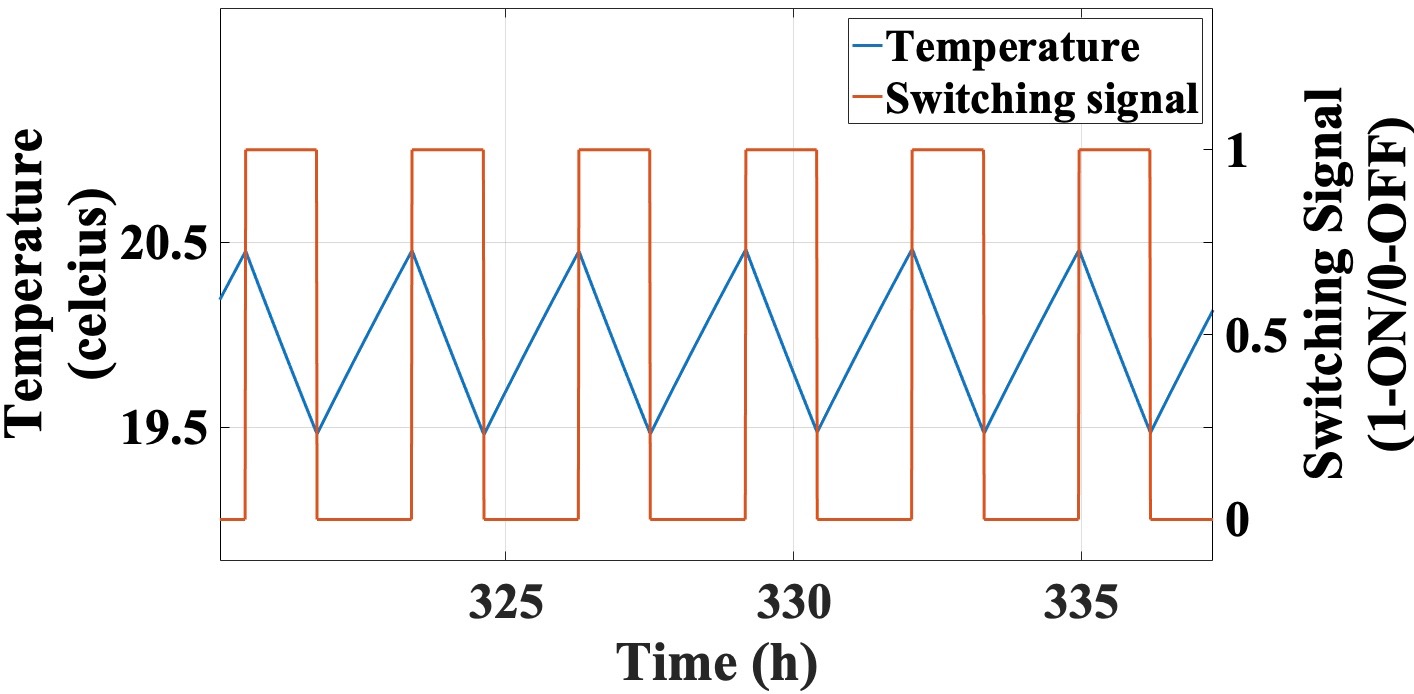}\\
(a)\\
\includegraphics[scale=0.3]{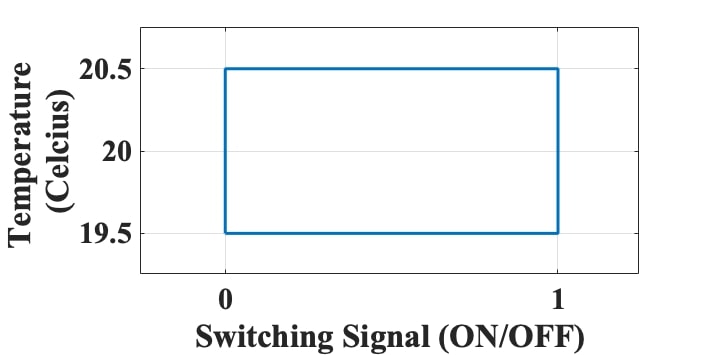}\\
(b)\\
\end{tabular}
\caption {Hybrid model. $\left(a\right)$ Temperature variation of a single TCL with respect to time. $\left(b\right)$ Phase portrait of a TCL system dynamics.}
\label{fig:n_tcl}
\end{figure}

\begin{equation}
\begin{aligned}
 \dot{T}(t)&=-\frac{1}{RC}[T(t)-T_a+s(t)PR],\\
 s(t)&= 
\begin{cases}
0 & \text{if} \ T(t) < T_{min},\\
1 & \text{if} \ T(t) > T_{max},\\
s(t) & \text{otherwise},
\end{cases}
 \label{eq1}
\end{aligned}
\end{equation}

\begin{table}[b!]
\begin{center}
\caption{List of parameters}\label{tb:methods}
\begin{tabular}{cccc}
Parameter & Meaning & Value & Unit\\\hline
$T_{a}$ & Ambient temperature & $32$ & $^{\circ}$C\\
$\delta$ & Thermostat deadband & $1$ & $^{\circ}$C\\
R & Thermal resistance & $2$ & $^{\circ}$C/kW\\
C & Thermal capacitance & $10$ & kWh/$^{\circ}$C\\
P & Energy transfer rate & $14$ & kW\\
K & Coupling strength & $0.267$ & Mhz\\
\hline
\label{tab1}
\end{tabular}
\end{center} 
% Reproduced with permission from Pratik Bajaria and N. M. Singh, Sustainable Energy, Grids, and Networks, Volume 18, June 2019. Copyright 2019  Elsevier B.V.
\end{table}

where $T_a$ and $P$ are ambient temperature and average energy transfer rate of a TCL respectively. The minimum and maximum temperature of TCLs are $T_{min}=\left(T_s-\delta/2\right)$ and $T_{max}=\left (T_s+\delta/2 \right)$ respectively, $T_s$ and $\delta$ being the thermostat set point temperature and deadband respectively. $s(t)$ is the switching signal of the TCLs. FIG. \ref{fig:n_tcl}$\left(a\right)$ depicts the temperature dynamics where the TCLs oscillates between $19.5^\circ C$ and $20.5^\circ C$ while FIG. \ref{fig:n_tcl}$\left( b \right)$ is phase portrait of the TCL obtained using hybrid model \eqref{eq1} and data from TABLE \ref{tab1}.
\par Thus, for $N$ TCLs, the aggregate power consumption is given by,

\begin{equation}
\begin{aligned}
 P_ {agg} =\sum^{N}_{j=1} P\eta_j s_j(t),
  \end{aligned}
  \label{eq2}
\end{equation}

where $\eta_j$ is the coefficient of performance of $j$th TCL. For simplicity of calculation, we consider $\eta_j=\eta=1$ for all TCLs.

\begin{rem}
All the parameters used henceforth are taken from TABLE \ref{tab1} unless otherwise specified.
\end{rem}

\begin{figure}[t!]
\centering
\begin{tabular}{c}
\includegraphics[scale=0.46]{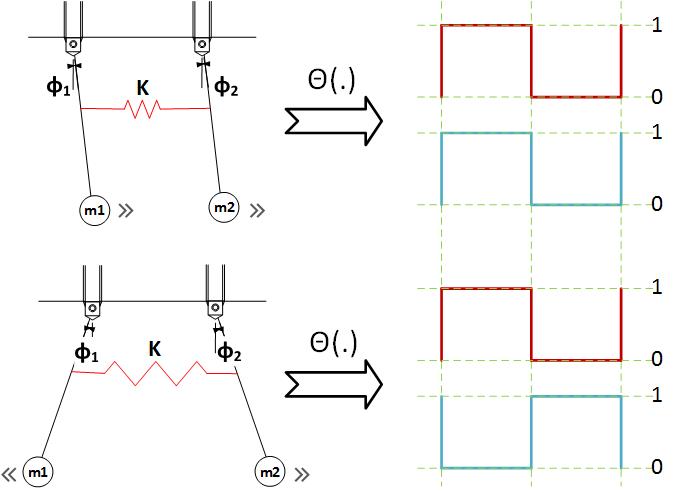}
\end{tabular}
\caption{(From top to bottom) Pendulums coupled by a spring, oscillating in two equilibrium modes; Oscillations in in-phase mode (Top), Oscillations in anti-phase mode (Bottom).}
\label{fig:pend_perspective}
\end{figure}
\section{Motivation} \label{sec3}
\subsection{Coupled Pendulums}\label{subsec1}

To further understand the effect of TCL switching oscillations on aggregate power consumption, we correlate the periodic behavior of TCL switching with a setup of spring coupled pendulums. It is well known that the Kuramoto model can be well understood using a system of coupled oscillators, and various modes of oscillations can be analyzed. Consider two pendulums coupled with a spring, as shown in FIG. \ref{fig:pend_perspective}, where $m_i$ are masses, $\phi_1$ and $\phi_2$ the phases and $K$ is the spring constant. For the sake of simplicity, assume both the pendulums to be identical in terms of system parameters and dissipative losses to be zero. 
As known, the phase of coupled pendulums exhibit `in-phase' (phase difference $=0$ radians), `anti-phase' (phase difference $=\pi$ radians) and `quasi in-phase' ($0<$ phase difference $<\pi$ radians) type oscillatory behavior \cite{strogatz2008}. Let all the phases measured in the anti-clockwise direction be positive and assume a constant external force that drives these oscillations. These external forces can be impulsive or persistent in time. It can be seen analogous to the addition of an external delay to maintain required steady-state behavior (see FIG. \ref{fig:pend_perspective}).

\par Now, all external forces (delay $=0$ radians) that keep aggregate switching amplitudes (a mathematical sum of phases) to its maximum can be considered to be oscillating in `in-phase' modes. All forces (delay $=\pi$ radians) that keep aggregate switching amplitudes to its minimum can be considered to be oscillating in `anti-phase' modes. All the other states oscillate between these two extrema can be termed as `quasi in-phase' modes. We portray `in-phase' and `anti-phase' oscillation modes pictorially, as shown in FIG. \ref{fig:pend_perspective}. Further, using the concept of clustering, we show how these modes come to benefit in achieving the required aggregate power.

\subsection{Boolean phase oscillator model for TCLs}\label{subsec 3}

\par In the previous work \cite{bajaria2019dynamic}, the author makes use of the Boolean phase oscillator model to incorporate the system dynamics and thereby control the aggregate power consumed ($P_{agg}$). Visualizing the TCLs as oscillators \cite{bajaria2019dynamic}, the author employs the Kuramoto framework to desynchronize TCLs. Beginning with the conventional Kuramoto equation \cite{bajaria2019dynamic}, to accommodate the digital switching signals, uses the Heaviside function, thereby imparting the Kuramoto framework the capability to couple digital signals. The standard Kuramoto equation can be expressed as follows,

\begin{equation}
 \begin{aligned}
 \dot{\phi}_i &= \omega_i + \frac{\epsilon}{N}\sum^{N}_{j=1,j\ne i} \left ( sin(\phi_j - \phi_i) \right ),\\
 \end{aligned}
 \label{eq3}
 \end{equation}
 where, $\omega_i$, $\phi_i$, $\epsilon$ and $N$ are natural frequency, phase of $i$th oscillator, gain and number of oscillators respectively, $\forall i \in \mathbb{Z^+}$.

\begin{figure}[t!]
\centering
\begin{tabular}{c}
\includegraphics[scale=0.25]{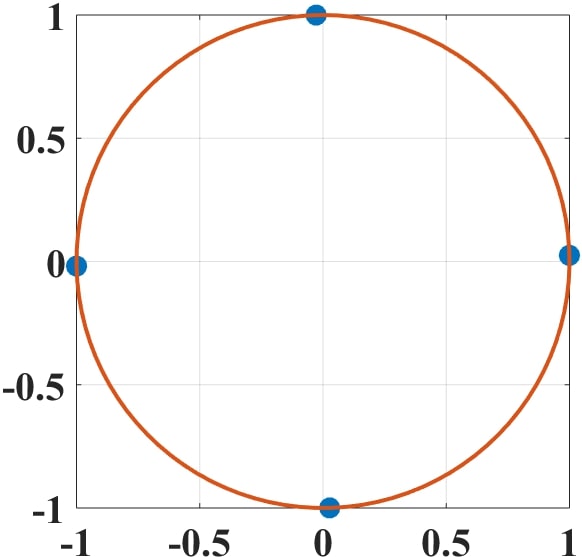}\\
(a)\\
\includegraphics[scale=0.16]{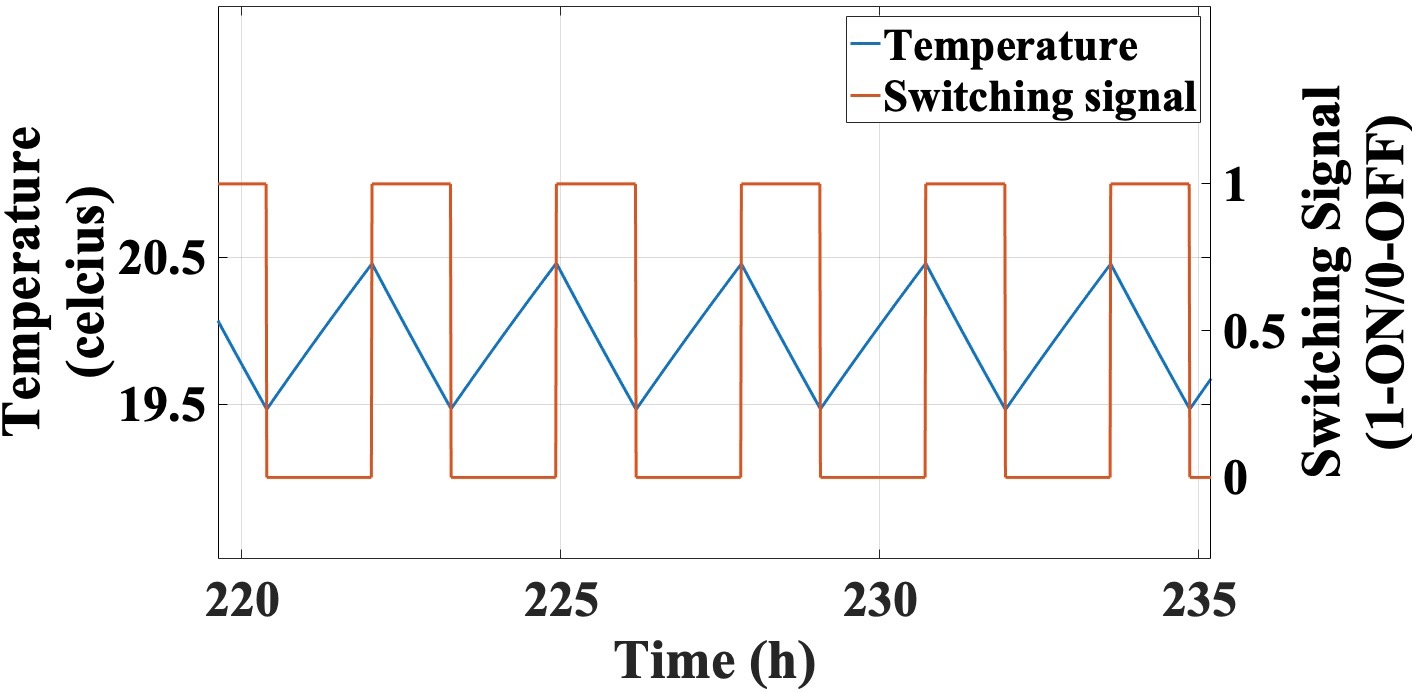}\\
(b)\\
\includegraphics[scale=0.18]{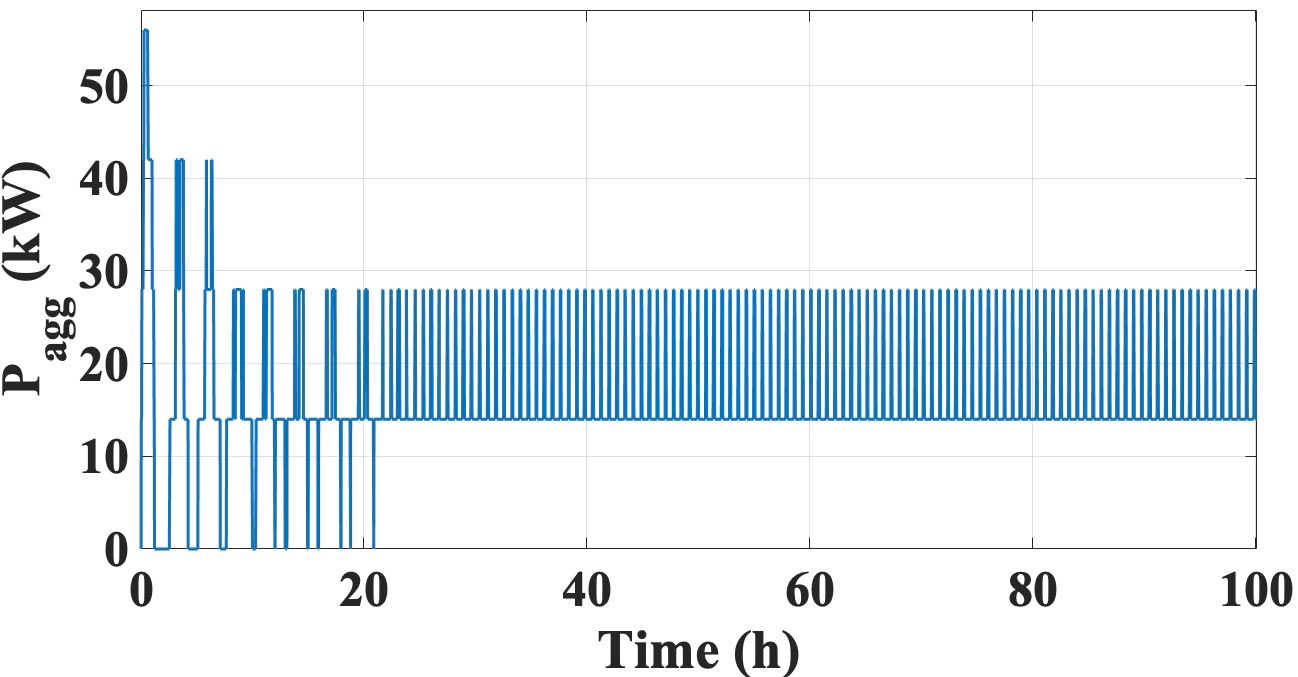}\\
(c)
\end{tabular}
\caption{Boolean phase oscillator model. (a) Angular separation for $N=4$ TCLs. (b) TCL temperature and switching signal dynamics. (c) $P_{agg}$ for $N=4$ TCLs at a given time.}
\label{fig:ensembletcl_dynamics}
\end{figure}

\par The inherent coupling dynamics of \eqref{eq3} enables the oscillators to establish communication between all TCLs. As mentioned earlier, the author using Heaviside function modifies \eqref{eq3} \cite{bajaria2019dynamic} and the resultant equation can be given as,
 \begin{equation}
 \begin{aligned}
 \dot{\phi}_i = \omega_{i}+K\sum^{N}_{j=1,j\ne i} \left | \Theta \left [ sin(\phi_j) \right ] - \Theta\left [ sin(\phi_i + \alpha_{ij}) \right ] \right |,\\
 \end{aligned}
 \label{eq4}
 \end{equation}  
 \par where $\Theta$ is Heaviside function given by,
  
 \begin{equation}
  \begin{aligned}
  \Theta[\sin{(.)}]= \begin{cases}
  \text{$0$} &\quad\text{$if$ $\sin{(.)}<0$},\\
  \text{$1$} &\quad\text{$if$ $\sin{(.)} \geq 0$},
  \end{cases}
  \end{aligned}
  \label{eq5}
 \end{equation}

 \par and $K$ is the coupling between oscillators. The phase separation $\alpha_{ij}$ in \eqref{eq4} induces a phase separation between the $i_{th}$ and $j_{th}$ oscillator, or in this case TCLs as shown in FIG. \ref{fig:ensembletcl_dynamics}$\left(a\right)$. Following equations characterize temperature dynamics and stabilization of aggregate power consumption as depicted in FIG. \ref{fig:ensembletcl_dynamics}$\left( b \right)$ and \ref{fig:ensembletcl_dynamics}$\left( c \right)$ respectively,

 \begin{equation}
  \begin{split}
 \dot{T}_i &= -\frac{1}{[RC]_{i}} \left [ T_i (t) - T_a + s_i(t)[PR]_{i} \right ],\\
 \dot{\phi}_i &= \omega_{i}+K\sum^{N}_{j=1,j\ne i} \left | \Theta \left [ sin(\phi_j) \right ] - \Theta\left [ sin(\phi_i + \alpha_{ij}) \right ] \right |,\\
 s_i(t) &=\Theta \left [ sin(\phi_i) - s_{i,0} \right ],\\
 s_{i,0} &= sin\left [ \frac{\pi - T_{ON,i}}{2}\right ],\\
 \end{split}
 \label{eq6}
 \end{equation}

Where $T_{ON,i}$ is ON time of the $i$th switching signal (mapped to its equivalent phase in radians), $\omega_i$ is the frequency of switching, $K$ is the coupling constant. A bias, ($s_{i,0}$), helps the generated signal map the duty cycle in the Kuramoto framework.

\begin{figure}[t!]
\centering
\begin{tabular}{c}
\includegraphics[scale=0.195]{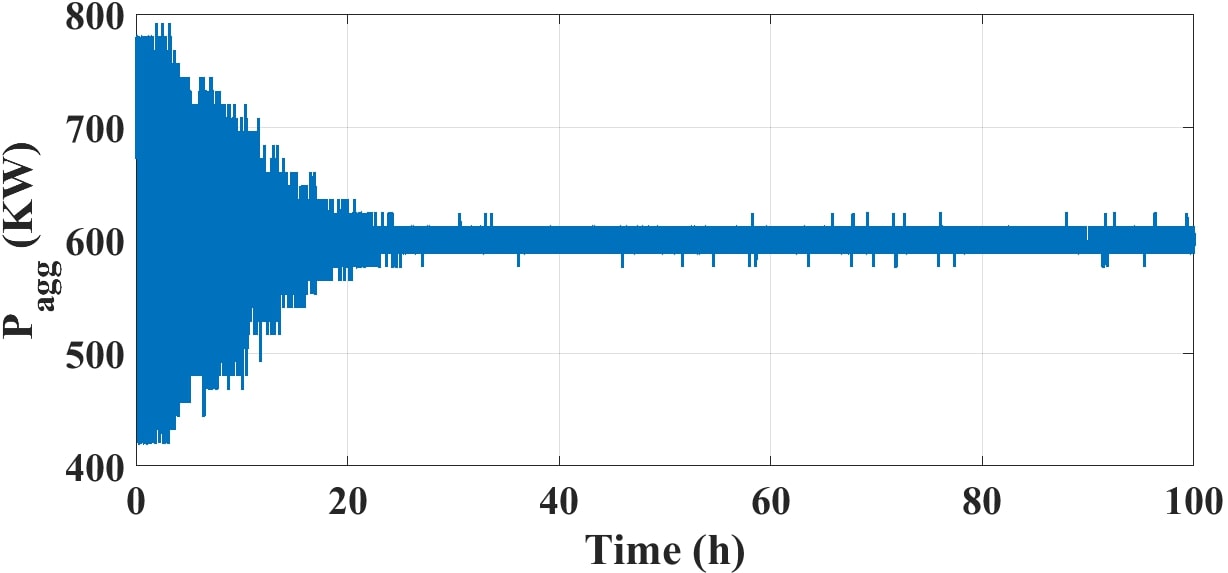}
\end{tabular}
\caption{Kuramoto based model. Aggregate power consumption of $N=100$ heterogeneous TCLs.}
\label{fig:n_100_ensembletcl}
\end{figure}

\par In the previous work \cite{bajaria2019dynamic}, the idea is further supported by simulations results. It was thus conclusively stated that the power oscillations and aggregation are minimum when the phase delay ($\alpha_{ij}$) is $2\pi/N$.

\par To delineate this theory, the author \cite{bajaria2019dynamic} takes a case of $N=4$ homogeneous TCLs, with the natural frequencies as $\omega_i = 0.55$Hz. Furthermore, the TCLs desynchronize at their respective phase differences with their common frequency being $\omega_{FFT} = 0.298$Hz, where $\omega_{FFT}$ is calculated using fast Fourier transform $\left(FFT\right)$ of the switching signal. Finally, the Kuramoto model desynchronizes TCLs at a common frequency lesser than their mean frequency. FIG. \ref{fig:ensembletcl_dynamics}$\left(c\right)$ shows the power vs. time plot for $N=4$ TCLs.

\par The author then takes a case of $N=100$ heterogeneous TCLs with their natural frequencies in the range of $\pm5\%$ of its nominal value. In FIG. \ref{fig:n_100_ensembletcl}, it is observed that the load fluctuations are minimized to within $\pm 2\%$ after applying Kuramoto based model.

\begin{rem}
Frequency calculated in the previous work \cite{bajaria2019dynamic} are scaled down by hours, i.e., $\omega_{i}$ is calculated as $1/T$ where, $T$ is in hours. However, in this paper we consider time in seconds.
\label{reamrk}
\end{rem}

\subsection{Shortcomings of Boolean phase oscillator model}\label{subsec 4}
\par Although the Boolean phase oscillator model \cite{bajaria2019dynamic} provides motivating results, some shortcomings in the model proposed can be listed as follows:

\begin{enumerate}

\item It must be noted that the natural frequency $\omega_i$ used in \eqref{eq6} can be easily calculated using mathematical manipulations as follows,

\begin{equation}
\begin{split}
  \omega_i = & \ \omega_{FFT}
  -K\sum^{N}_{j=1,j\neq i} \left | \Theta \left [ sin(\phi_j) \right ] - \Theta\left [ sin(\phi_i + \alpha_{ij}) \right ] \right|,
\end{split}
 \label{eq7}
\end{equation}

\par However, it is observed that as the number of TCLs increases, the value of $\omega_i$ needs correction. For instance, $N=16$ requires $\omega_i = 8.81 $ Hz and $N=100$ needs $\omega_i=62.18$ Hz, to achieve $\omega_{FFT} = 0.298$ Hz. Moreover, as the number of TCLs increases, $\omega_{i}$ computed from \eqref{eq7} requires an additional correction factor $e$ to give desired results. Failing to find the correct $e$ leads to erroneous deadband values. Also, since the system under consideration consists of mechanical devices such as compressors and switches, such high frequencies are undesirable. Additionally, the TCLs are frequency sensitive devices, and drastic changes in the frequency must be avoided to prolong the product life \cite{mehta2014}. In a practical scenario, where system parameters (e.g., $N, \alpha_{ij}$) change dynamically, computation of $\omega_i$ in real-time would seem difficult. Thus, rendering the Boolean phase oscillator model difficult to be practically realized.

\item $K$ is a parameter that is computed using dynamical equations governing the system parameters \cite{dorfler2013synchronization}. In case of TCLs, the value of $K$ can be calculated from the induced time delays in a programmable frame \cite{bajaria2019dynamic}. However, the inter-dependence of $K$ and $\omega_{i}$ in \eqref{eq6} adds complexity to its accurate computation.

\end{enumerate}
\begin{figure}[t!]
\centering
\begin{tabular}{c}
\includegraphics[scale=0.18]{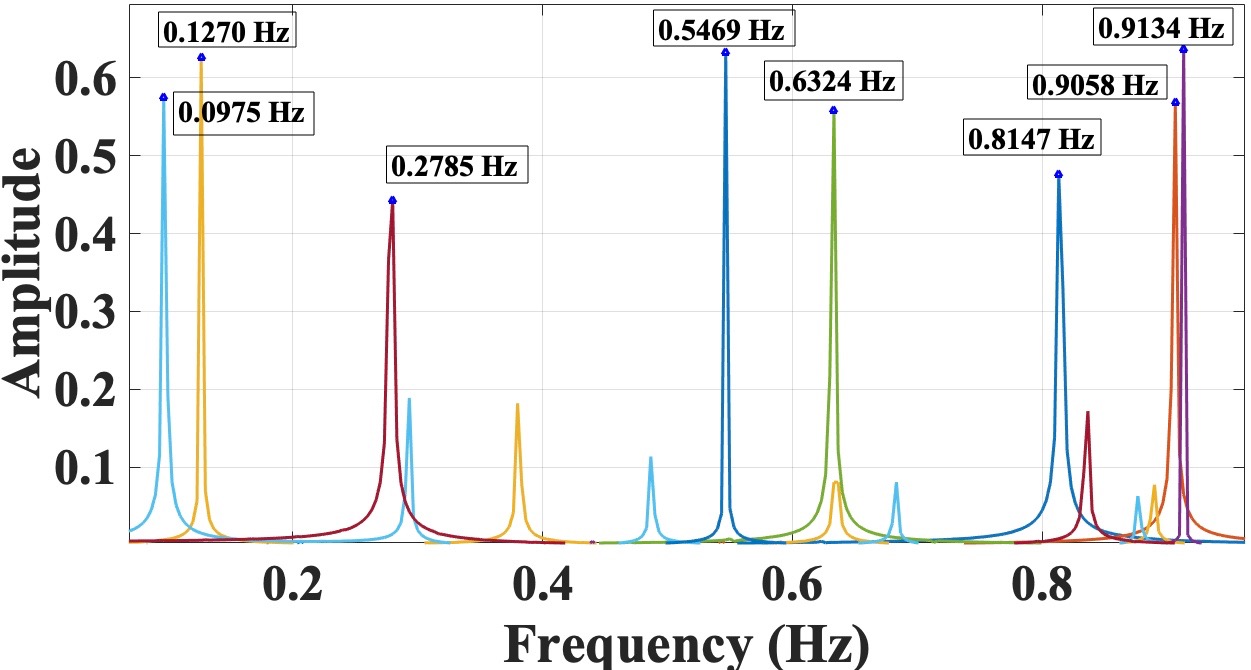}\\
(a)\\
\includegraphics[scale=0.18]{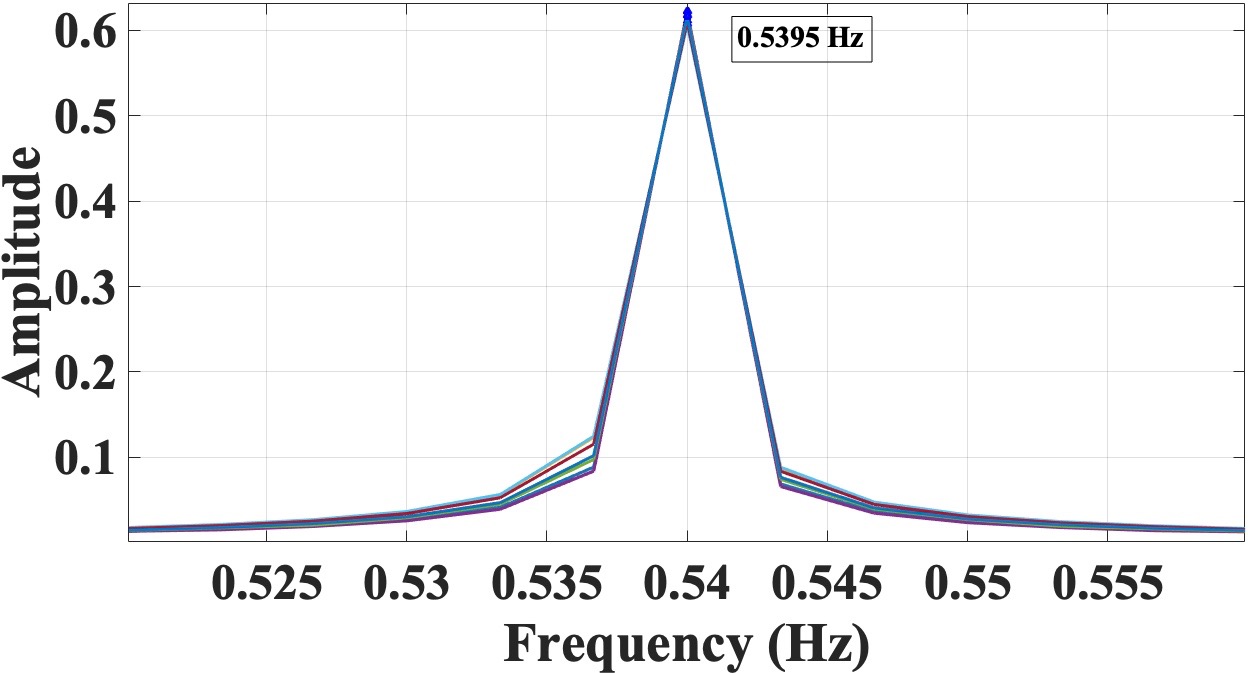}\\
(b)\\
\end{tabular}
\caption{Frequency dynamics. (a) FFT of switching signal at $t$ = $t_{0}$. (b) FFT of switching signal at $t$ = $t_{f}$.}
\label{fig:n_8_tcl}
\end{figure}

\section{Distributed Averaging Model for the TCLs}\label{sec 4}

\par Consider an undirected graph $G = \{V,E\}$ with set of nodes $N$ in $V$ and edge set $E$. Each edge of the graph $E \in G$ is an unordered pair of distinct nodes. A real scalar quantity $x_i$(t) is associated with node $i$ at any given time $t$. The distributed averaging consensus algorithm computes the average $\frac{1}{N}\sum ^N_{i=1} x_i(0) $ iteratively thus, facilitating a local communication link between the nodes. The nodes are updated continuously based on their states as well as the states of all other neighboring nodes.

\par A widely accepted and used linear iterative algorithm can be stated as follows,
\begin{equation}
 \begin{aligned}
  \dot{x}_i(t) &= \sum ^ N_{j=1, j\neq i}W_{ij}\Big(x_j(t)-x_i(t)\Big),
 \end{aligned}
 \label{eq8}
\end{equation}

for $i \in \{ 1,\ldots,n \}$ and $t \in \mathbb{R^+}$ and $W_{ij}$ being the weight associated with each edge $E_{ij}$. Since the nodes, in the presented case, are undirected, the weights associated with them are symmetric i.e., $W_{ij} = W_{ji}$.
\begin{equation}
 \begin{aligned}
  W = w_{ij}(\mathbbm{1}_{N\times N} - diag(\mathbbm{1}_{N})),
 \end{aligned}
 \label{eq9}
\end{equation}

where, $w_{ij}$ is the weight of edge $E_{ij}$ and $diag(\mathbbm{1}_{N})$ is diagonal matrix with $\mathbbm{1}_{N}, \mathbbm{1}_{N\times N}$ denoting a vector and matrix of ones, respectively. Thus, the matrix $W$ is the communication matrix that describes the strength of the link between two nodes. Analogous to frequency synchronization seen in the Kuramoto framework \eqref{eq6} we extend the distributed theory to the system of TCLs. The nodes are similarized to TCLs and the scalar quantity to the frequency of switching signal as follows,

\begin{equation}
\begin{split}
 \dot{f_{i}} &= W\sum^{N}_{j=1,j \neq i} (f_{j} - f_{i}),\\
\end{split}
\label{eq10}
\end{equation}

where $W$ is the tuning parameter of synchronization, left to the control designer. \eqref{eq10} synchronizes the frequencies randomly chosen from distribution to their mean value, as shown in FIG. \ref{fig:n_8_tcl}. Thus, \eqref{eq6} can be re-framed as follows,
 \begin{equation}
  \begin{split}
 \dot{T}_i &= -\frac{1}{[RC]_{i}} \left [ T_i (t) - T_a + s_i(t)[PR]_{i} \right ],\\
 \dot{f_{i}} &= W\sum^{N}_{j=1,j \neq i} (f_{j} - f_{i}),\\
 s_i(t) &=\Theta \left [ sin(2\pi f_{i}t + \alpha_{ij}) - s_{i,0} \right ],\\
 s_{i,0} &= sin\left [ \frac{\pi - T_{ON,i}}{2}\right ].\\
 \end{split}
 \label{eq11}
 \end{equation}  
  
 \par It should be noted that \eqref{eq11} bifurcates the onus of controlling the frequency and angular separation.

\begin{figure}[t!]
\centering
\begin{tabular}{c}
\includegraphics[scale=0.18]{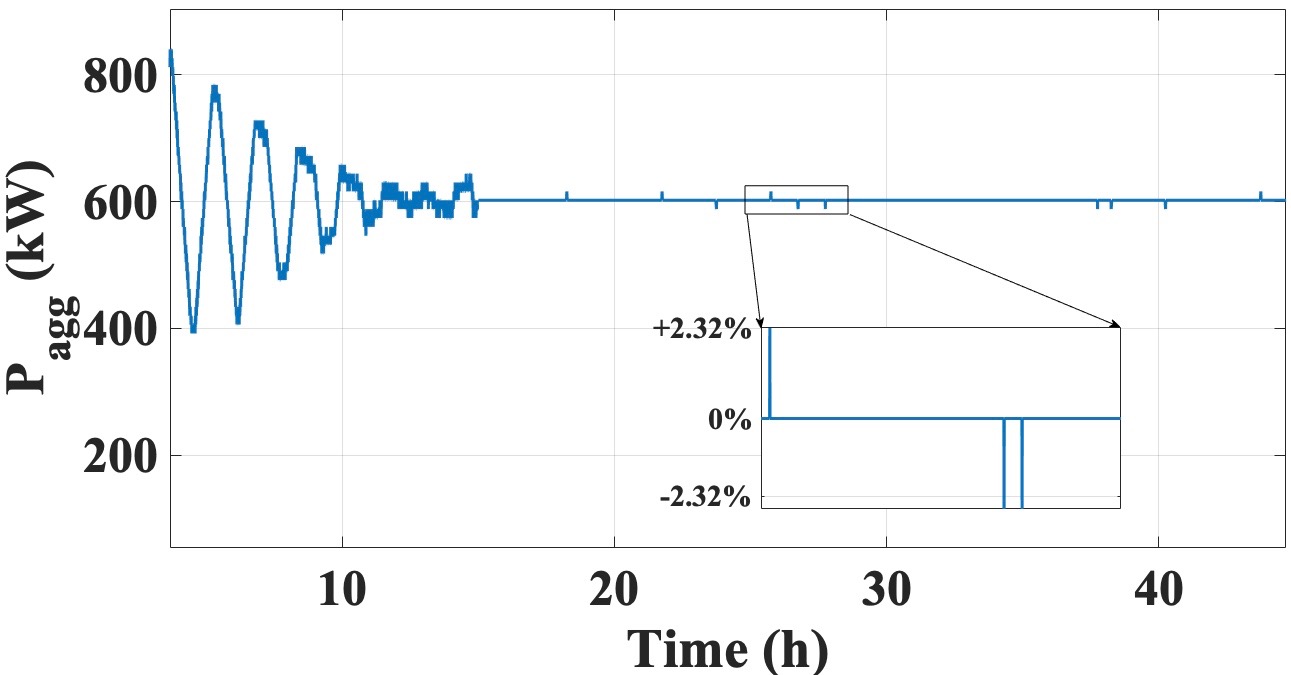}
\end{tabular}
\caption{$P_{agg}$ vs time showing the aggregate power consumed at a given time for $N=100$ heterogeneous TCLs.}
\label{fig:n_100_tcl_d}
\end{figure}

\par For the matter of comparison, a similar ensemble of heterogenous TCLs, as in FIG. \ref{fig:n_100_ensembletcl} (natural frequencies in the same range as of $\pm 5\%$ of $\omega_i = 0.271$Hz), are taken. Additionally, the value of weight($W$) chosen as $0.06$. By using \eqref{eq11}, the system desynchronizes, and FIG. \ref{fig:n_100_tcl_d} depicts the results obtained. Similar results can be noted by comparing the two plots, however, as opposed to Kuramoto based model in FIG. \ref{fig:n_100_tcl_d}, high frequency switching and computational hindrance have been avoided.

\section{Software and Hardware Implementation } \label{sec 5}

\par This section presents the application of the distributed averaging protocol in the form of software simulations and later describes its practical implementation.

\begin{table}[b!]
\begin{center}
\caption{List of Parameters (at set point temperature = $24^{\circ} C$)}\label{tb:tab2}
\begin{tabular} { cccc}
Parameter & Meaning & Value & Unit\\\hline
$f$ & frequency of TCL & 0.0027 & Hz \\
$\delta$ & thermostat deadband & 2 & $^{\circ}C$\\ 
$d$ & duty cycle of TCL & 0.5083 & - \\ 
\hline
\end{tabular}
\end{center}
\end{table}

\begin{table}[b!]
\begin{center}
\caption{List of Parameters (at set point temperature = $27^{\circ} C$)}\label{tb:tab3}
\begin{tabular} { cccc}
Parameter & Meaning & Value & Unit\\\hline
$f$ & frequency of TCL & 0.0031 & Hz \\
$\delta$ & thermostat deadband & 3 & $^{\circ}C$\\ 
$d$ & duty cycle of TCL & 0.452 & - \\  
\hline
\end{tabular}
\end{center}
\end{table}

\subsection{Case Study}\label{subsec 1}

\par To make the two cases put forth as realistic as possible, real time data of a TCL has been used to simulate the cases. For this, a TCL of a certain make was observed and the data listed in the TABLE \ref{tb:tab2} and TABLE \ref{tb:tab3} were collected. This TCL had a power rating of $1.66$kW and capacity of $1.5$tons. The values $R=9.8 ^{\circ}$C/kW and $C=0.0746$kW/$^{\circ}$C were computed using \eqref{eq1}, TABLE \ref{tb:tab2} and TABLE \ref{tb:tab3}.

\begin{figure}[t!]
\centering
\includegraphics[scale=0.18]{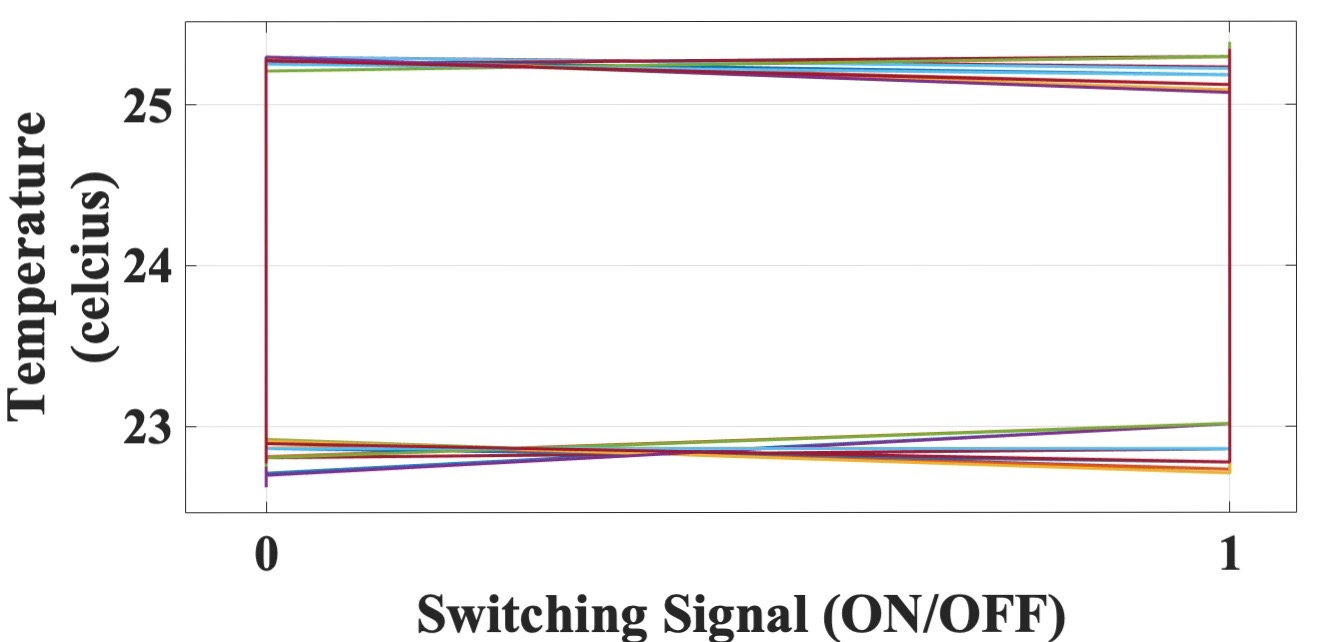}\\
\caption{Phase portrait for $N=10000$ TCLs.}
\label{fig:phaseportraitN100}
\end{figure}

\begin{enumerate}
 \item Case 1: $N=1000$
 \par For $N=1000$, heterogeneous TCLs are defined having a duty cycle in the range of $0.422$ to $0.482$, $\delta$ as $3.0^{\circ}C$ and $f$ between $0.0029$Hz and $0.0033$Hz respectively. The TCLs have been set to work at a set point temperature of $27^{\circ}C$.

\par Following the distributed averaging model, TCLs settle at their respective phase difference, i.e., $\alpha_{ij} = \alpha = 2\pi/N$, which in this case is $\alpha = \pi/500$. As depicted in FIG. \ref{fig:power_hetero_tcl_1}, the power fluctuation for $N=1000$ heterogeneous TCLs dampen over time and move towards steady state. The aggregate power at which the TCLs settle is $750.4472$kW and fluctuations in the power are $\pm2.67\%$. 
%750.3203KW mean value and 750.4472KW rms value.

\begin{figure}[t!]
\centering
\includegraphics[scale=0.18]{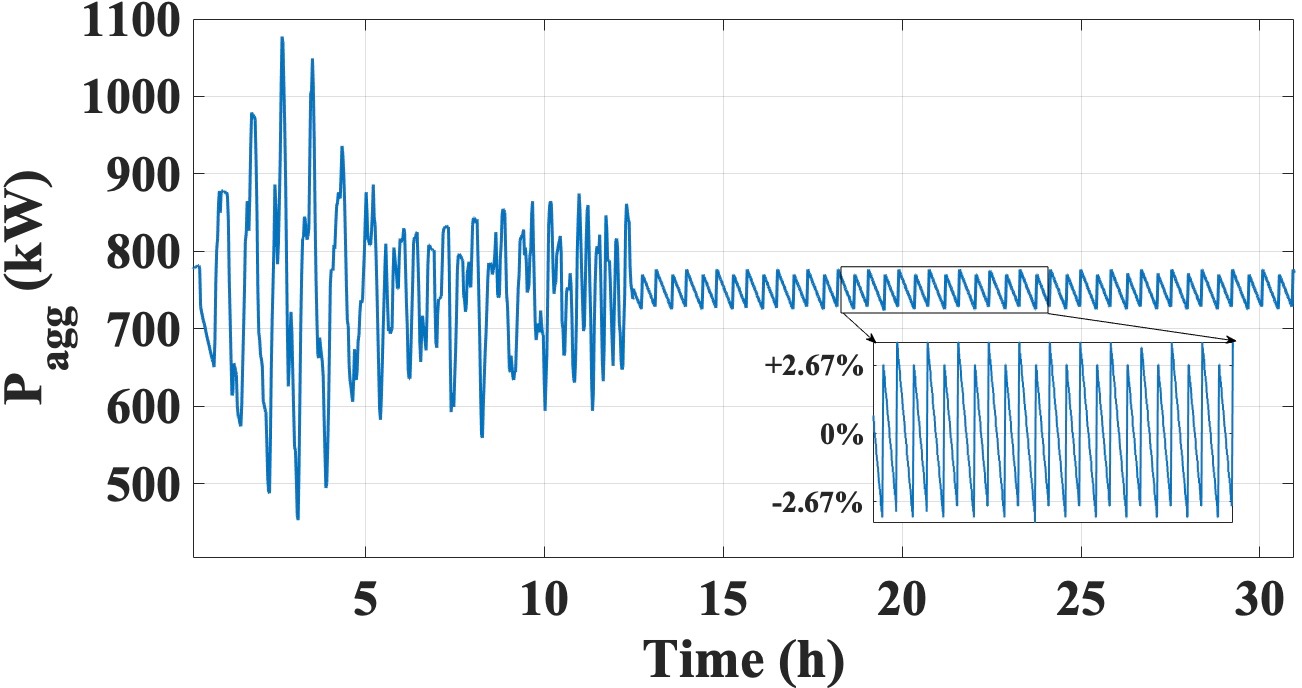}\\
\caption{$P_{agg}$ vs time showing the aggregate power consumed at a given time for $N=1000$ heterogeneous TCLs.}
\label{fig:power_hetero_tcl_1}
\end{figure}

\begin{figure}[t!]
\centering
\includegraphics[scale=0.19]{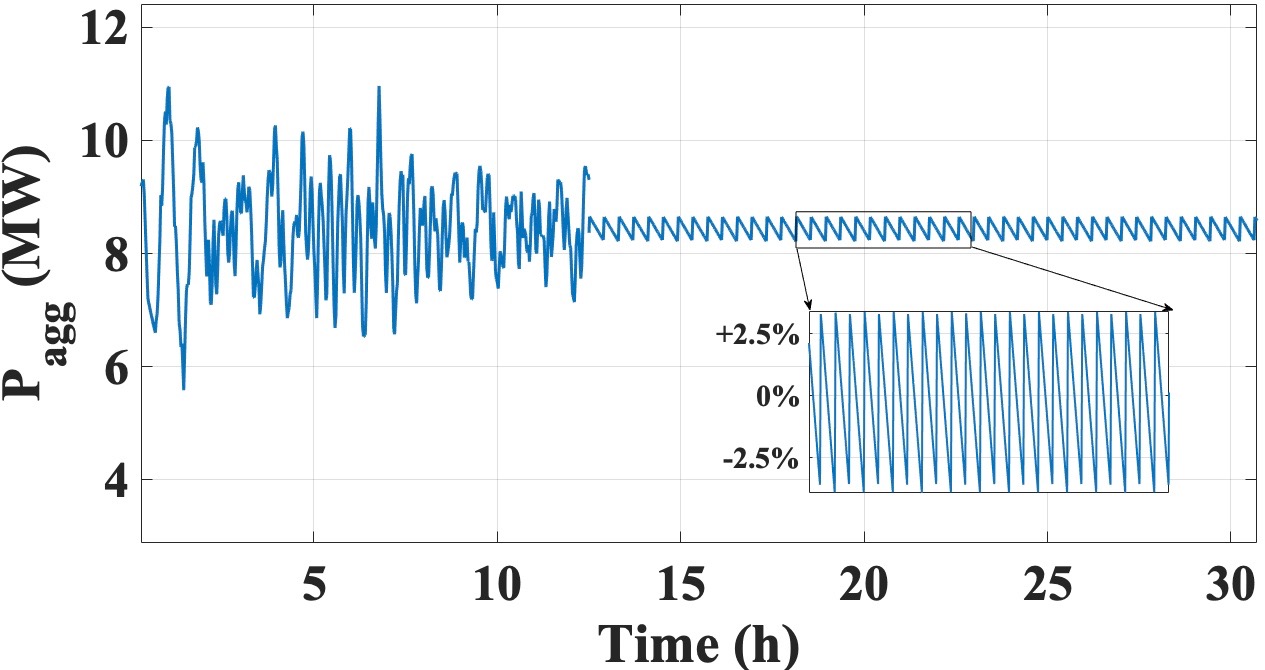}\\
\caption{$P_{agg}$ vs time showing the aggregate power consumed at a given time for $N=10000$ heterogeneous TCLs. }
\label{fig:power_hetero_tcl_2}
\end{figure}

\item Case 2: $N=10000$

\par For $N=10000$, heterogeneous TCLs are defined having a duty cycle in the range of $0.4812$ to $0.5354$, $\delta$ as $2.0^{\circ}C$ and $f$ between $0.0026$Hz and $0.0036$Hz respectively. The TCLs have been set to work at a set point temperature of $24^{\circ}C$.

\par Following the distributed averaging model, TCLs settle at their respective phase difference, i.e., $\alpha_i = \alpha = 2\pi/N$, which in this case is $\alpha = \pi/5000$. As depicted in FIG. \ref{fig:power_hetero_tcl_2}, the power fluctuation for $N=10000$ heterogeneous TCLs dampen over time and move towards steady state. The aggregate power at which the TCLs settle is $8.4453$MW, and fluctuations in power are $\pm2.5\%$. In FIG. \ref{fig:phaseportraitN100}, phase portraits of $100$ TCLs chosen randomly from the population are shown. It is clear that the portrait dissimilarity is minimal, although setpoint temperatures remain unaffected. %8.4444MW mean value and 8.4453MW RMS value
 \end{enumerate}

\begin{figure}[t!]
\centering
\includegraphics[scale=0.18]{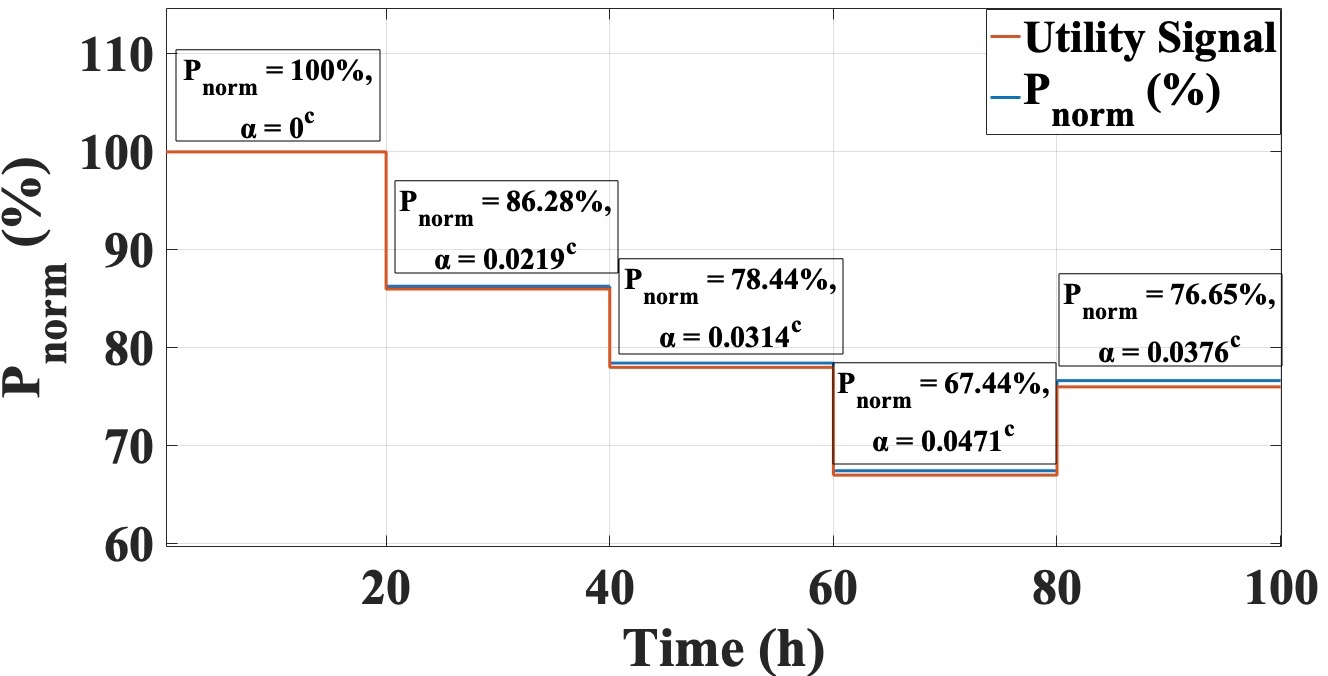}\\
(a)\\
\includegraphics[scale=0.18]{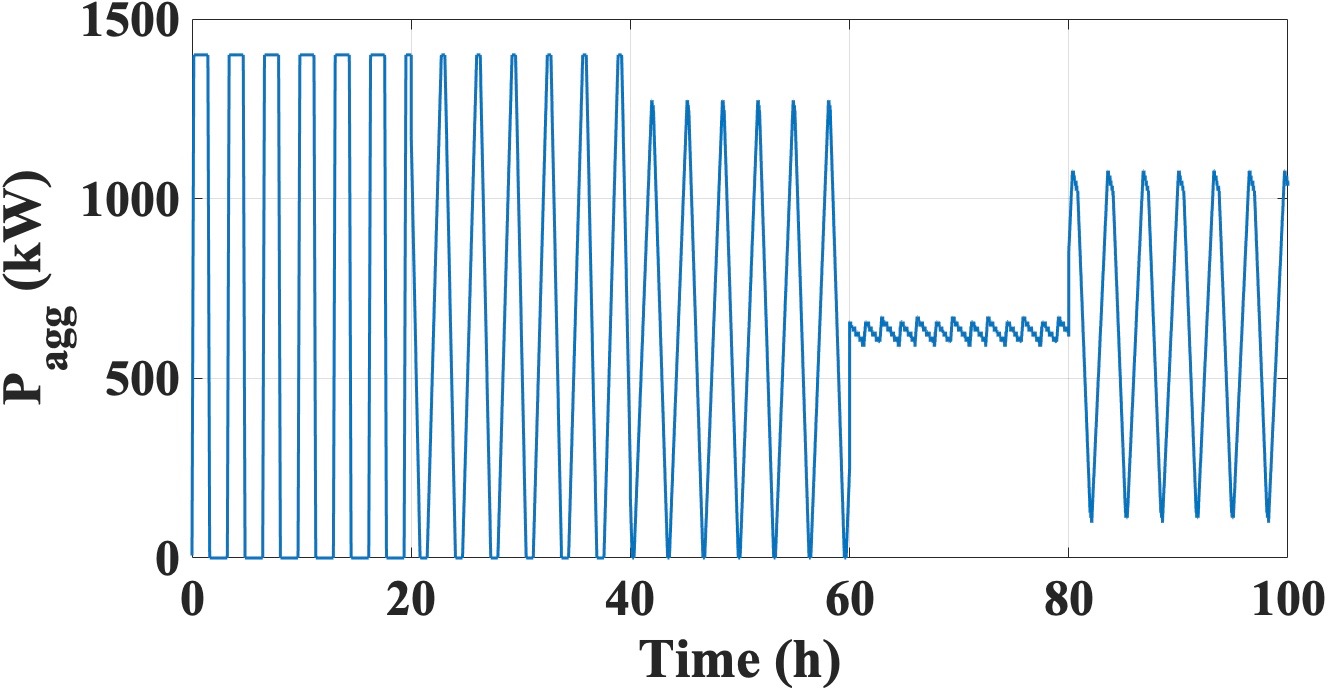}\\
(b)\\
\caption{(a) $P_{norm}(\%)$ vs time showing variation of $P_{norm}(\%)$ with $\alpha$. (b) $P_{agg}$ vs time plot showing variation of $P_{agg}$ with $\alpha$. }
\label{fig:prms_delay_Calculator}
\end{figure}

\subsection{Load Following}
\par Next, the authors show how a load-following case can be emulated, providing a valuable AS to the utility. In the previous work \cite{bajaria2019dynamic}, a function mapping RMS (root mean square) aggregate power of a population of TCLs with the effective delay induced was proposed. Using the same idea, we choose different utility demands, and the corresponding delay of $\alpha_{ij}$ is calculated. The effective power so obtained is normalized using total RMS power capacities; i.e.,

\begin{equation}
 \begin{aligned}
  P_{norm}(\%) = \frac{(P_{rms,agg}-P_{rms,\alpha})}{P_{rms,agg}}\times100,
 \end{aligned}
 \label{P_norm}
\end{equation}

\par where $P_{rms,agg}$ is the RMS of aggregate power, $P_{rms,\alpha}$ is the RMS of aggregate power at steady-state at $\alpha_{ij}=\alpha$ and $P_{norm}$ is the normalized power of all the TCLs along a given period respectively. These results can be seen from FIG. \ref{fig:prms_delay_Calculator}$\left(a\right)$ where utility demands for random reduction/increase in the total allowable aggregate power and the corresponding change in the delay induced causes controlled desynchronization and hence the required load following scenario is achieved. As seen from FIG. \ref{fig:prms_delay_Calculator}$\left(b\right)$($N=100, P=14kW$) it can be inferred that the resultant aggregate power is a significant AS to the centralized power grid.

\begin{figure}[t!]
\centering
\includegraphics[scale=0.18]{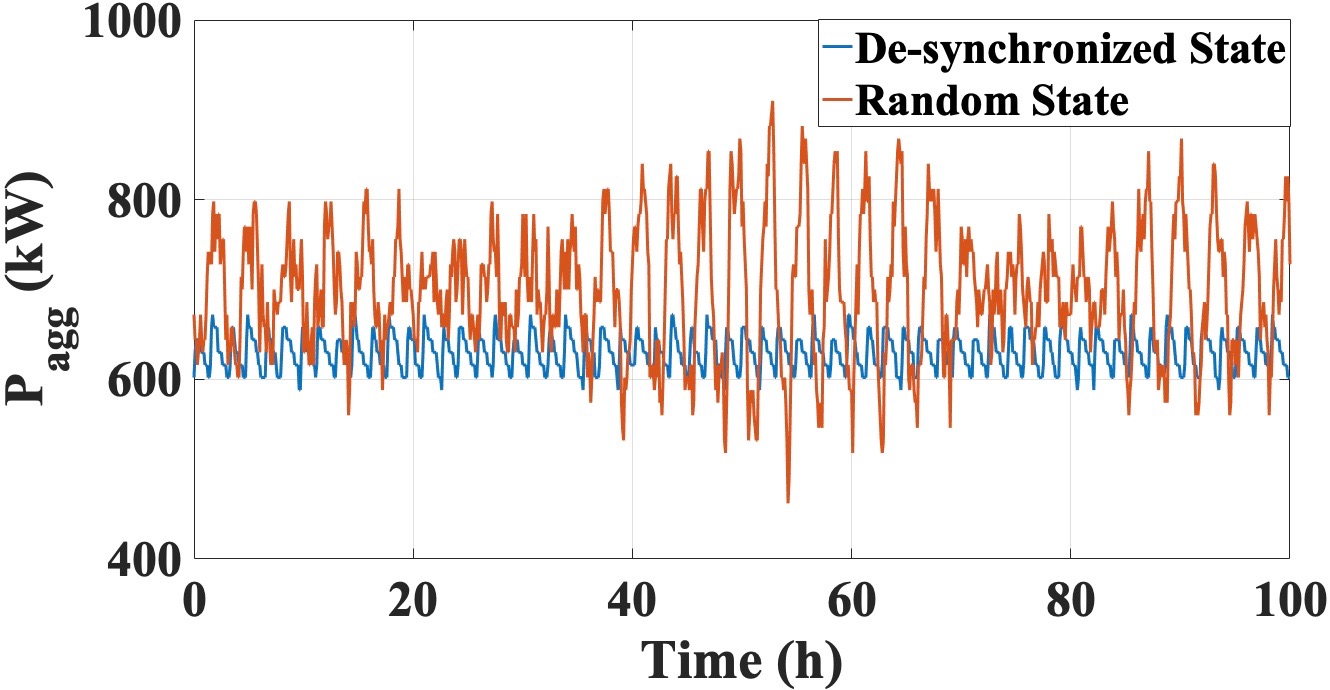}\\
(a)\\
\includegraphics[scale=0.185]{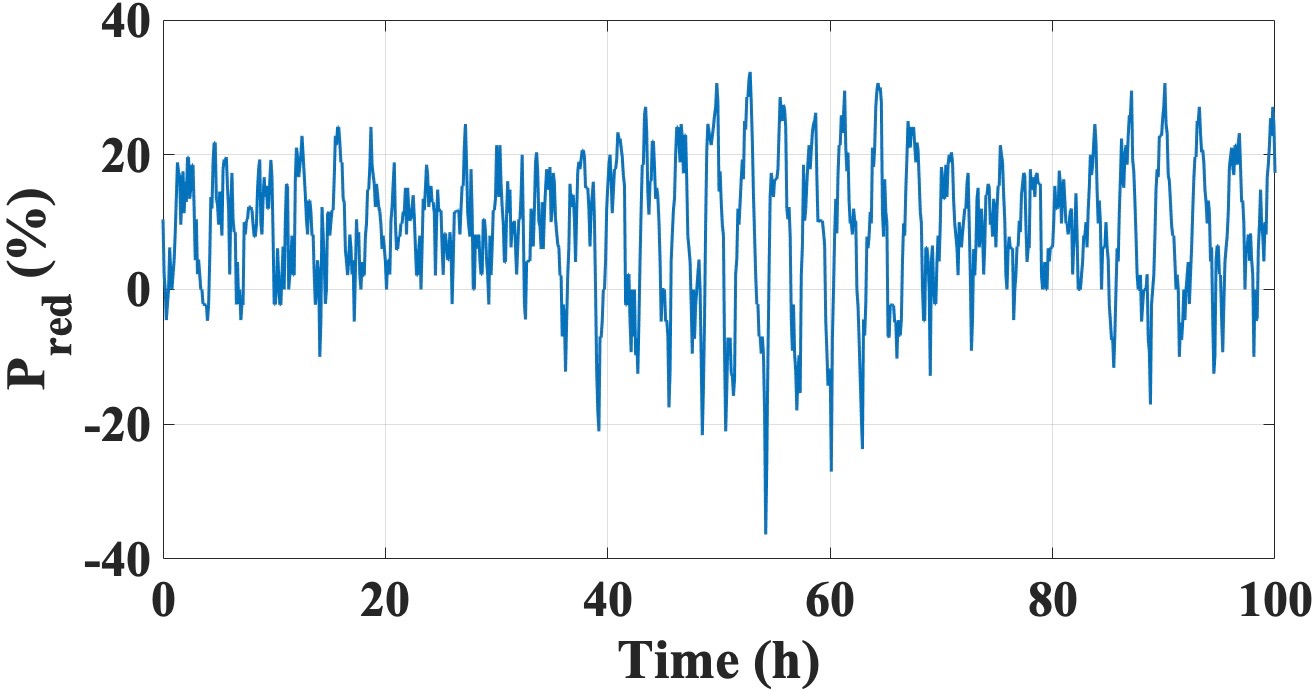}\\
(b)\\
\includegraphics[scale=0.18]{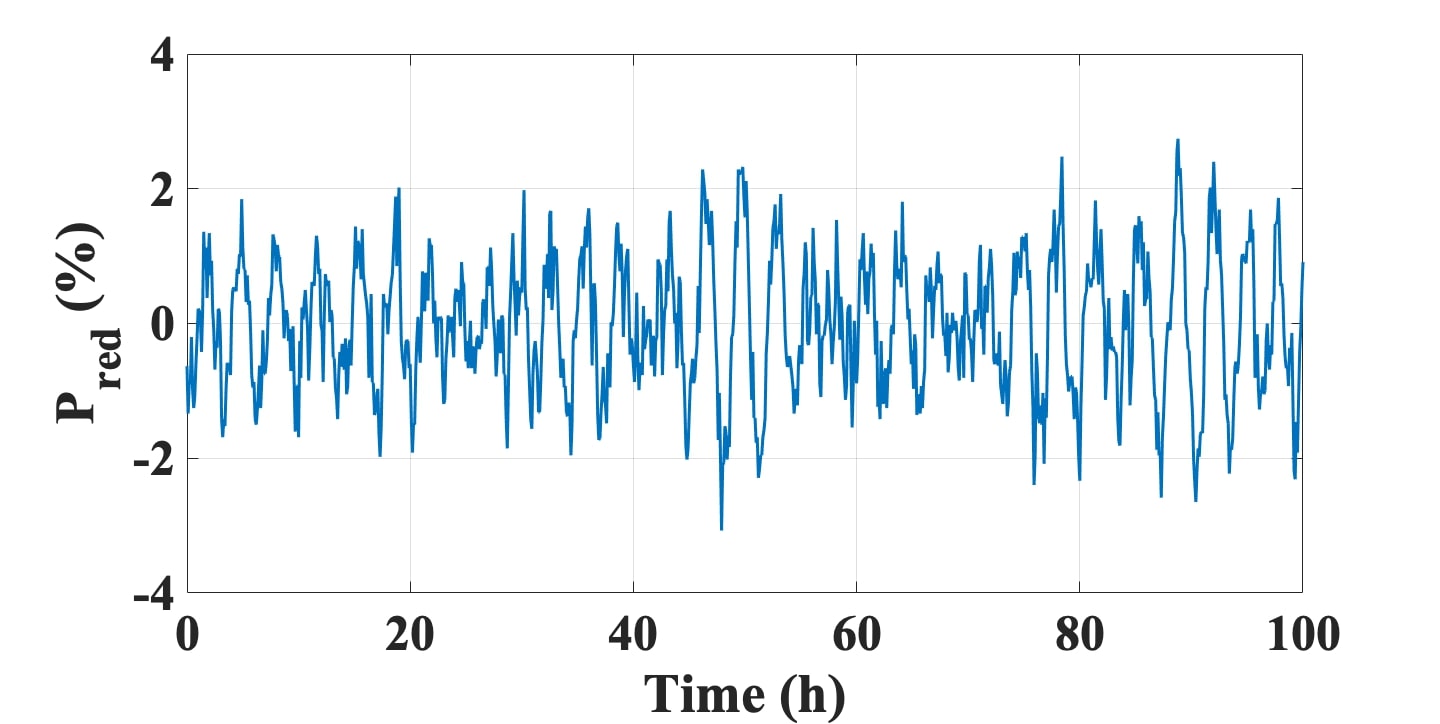}\\
(c)\\
\caption{(a) $P_{agg}$ vs time plot for random and desynchronized state for $N=100$. (b) $P_{red}(\%)$ vs time plot showing the reduction in fluctuations for $N=100$. (c)$P_{red}(\%)$ vs time plot showing the reduction in fluctuations for $N=10000$.}
\label{fig:p_fluctuations}
\end{figure}

\subsection{Effects in a Power Grid}
\par It should be noted that apart from the control of power aggregation, the proposed algorithm benefits the grid by reducing the power system fluctuations. To compare/visualize the same, we take a small population of $N=100$ TCLs. To exhibit random behavior i.e., depicting as-is characteristics without applying control), the frequency and phases of TCLs are chosen randomly from a distribution. Next, the same set of heterogeneous TCLs is acted upon by the distributed averaging protocol achieving effective desynchronization. As shown in FIG. \ref{fig:p_fluctuations}$\left(a\right)$, the system without application of distributed averaging strategy induce higher oscillations in the grid as compared to after application of the proposed scheme. In FIG. \ref{fig:p_fluctuations}$\left(b\right)$ the quantum of reduction is observed to be $\sim 40\%$ ($560 kW$) of random case, showing a significant reduction($P_{red}(\%)$) where,

\begin{equation}
 \begin{aligned}
  P_{red}(\%) = \frac{(P_{random}-P_{desynchronized})}{P_{random}}\times100.
 \end{aligned}
 \label{P_norm}
\end{equation}

Similarly, a case of $N=10000$ TCLs is also shown in FIG. \ref{fig:p_fluctuations}$\left(c\right)$. It can be seen a reduction of $\sim 2\%$ can be noticed which translates to $2.8MW$.
Hence, apart from providing load following objectives, the proposed scheme can also help reduce the amplitude of power system oscillations.

\subsection{Hardware Implementation}\label{subsec2}
\par
\begin{figure}[t!]
\centering
\includegraphics[scale=0.18]{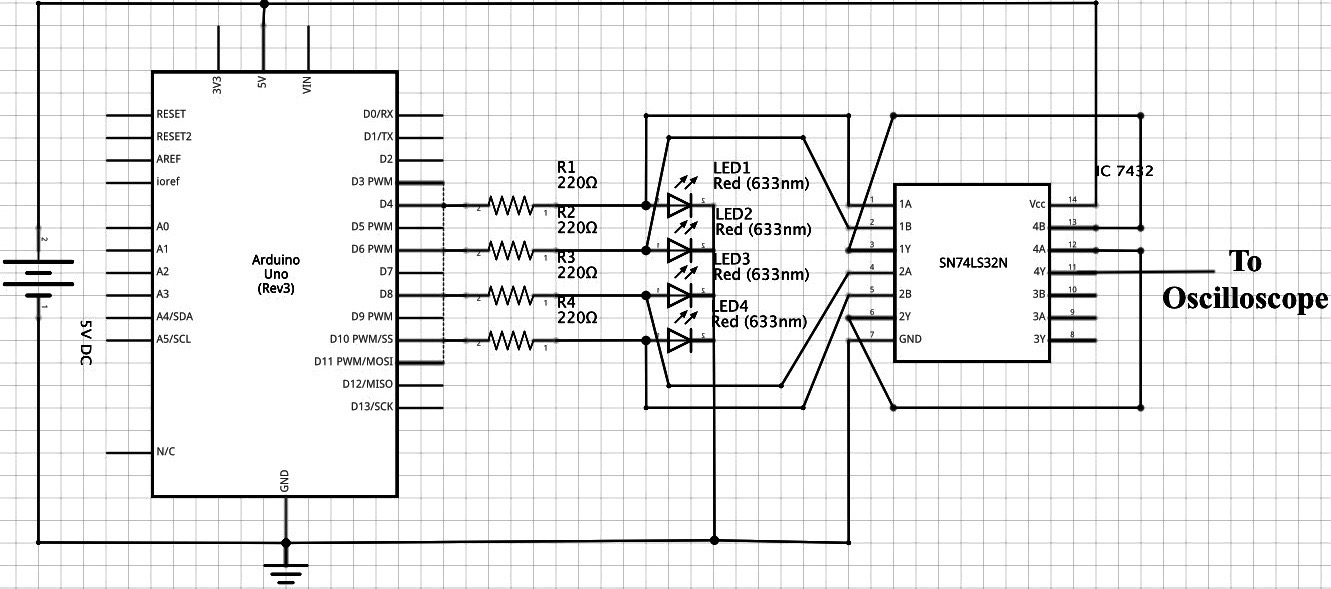}\\
\caption{Circuit diagram used for practical implementation.}
\label{fig:ckt_tcl_1}
\end{figure}

\begin{figure}[t!]
\centering
\includegraphics[scale=0.07]{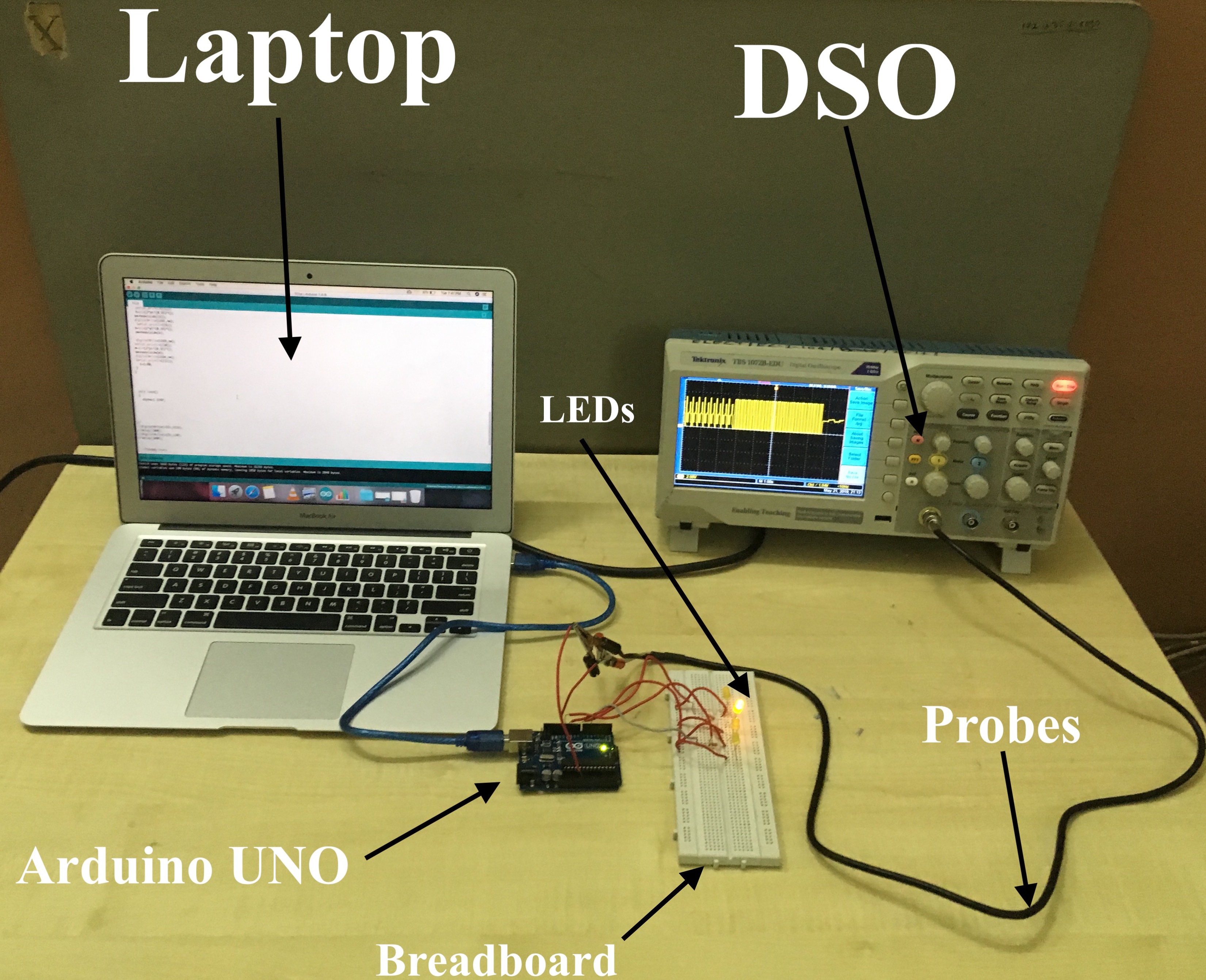}
\caption{Practical setup used for implementation of distributed averaging model.}
\label{fig:hardware_tcl_1}
\end{figure}

 To validate the presented theory, the authors have implemented the model on a hardware system. The authors use the circuit diagram as shown in FIG. \ref{fig:ckt_tcl_1}, which practically looks as seen in FIG. \ref{fig:hardware_tcl_1}.

\par Light-emitting diodes (LEDs) provide close resemblance to a switching device, analogous to TCLs. Hence, for the demonstration purpose, LEDs have been used in place of an actual TCLs. Since the number of LEDs was limited to four, a micro-controller with necessary computational capabilities and decent clock frequency was sufficient.

\subsubsection{Hardware}

\par The components used for the implementation include 4 LEDs, micro-controller, a breadboard, 5V DC supply, wires for connection, and probes to display the output on the oscilloscope. The digital storage oscilloscope (DSO) used has the following specifications,
\begin{itemize}
\item Product Range: TBS1000B-EDU
\item Scope Channel: $2$
\item Bandwidth : $70$MHz 
\item Sampling Rate :$1$ GSPS
\item Display Memory Depth : $2.5$kpts
\item Scope Display Type: WVGA LCD Colour
\item Plug Type: EU, SWISS, UK
\end{itemize}
The micro-controller used is Arduino Uno with following specifications,
\begin{itemize}
 \item Micro-controller: ATmega$328$
\item Operating Voltage: $5$V
\item Digital I/O Pins: $14$ (of which $6$ are PWM pins)
\item Analog Input Pins: $6$
\item DC Current per I/O Pin: $40$ mA
\item DC Current for $3.3$V Pin: $50$ mA
\item Clock Speed: $16$ MHz
\end{itemize}

\begin{figure}[t!]
\centering
\includegraphics[scale=0.3]{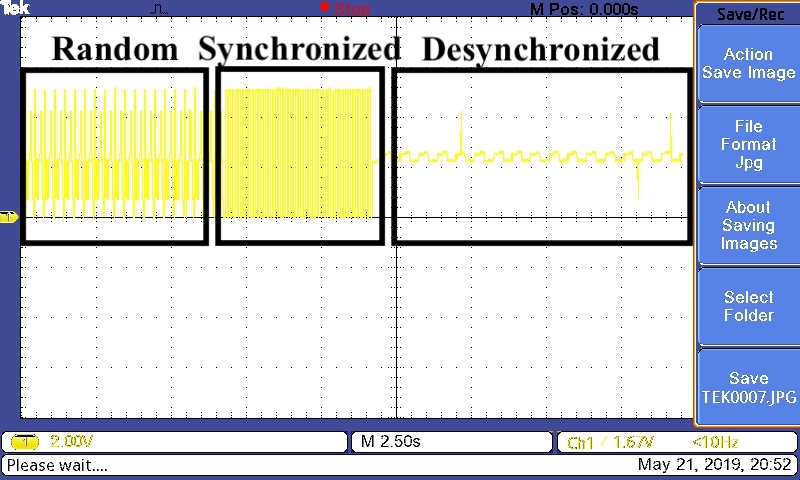}\\
\caption{Oscilloscope output for $N=4$ depicting random, synchronized and desynchronized period of LEDs.}
\label{fig:osco_tcl_1}
\end{figure}

\subsubsection{Implementation}

\par The setup for the practical implementation of the distributed averaging model is shown in the FIG. \ref{fig:hardware_tcl_1}. Based on the distributed averaging model, as described by \eqref{eq11}, the logic is coded and burned on Arduino. The Arduino generates switching signals for the LEDs. The output of the LEDs is observed on the oscilloscope.

\par As it can be seen, the output shown in FIG. \ref{fig:osco_tcl_1} is similar to that of FIG. \ref{fig:n_100_tcl_d}.

\par In the beginning, i.e., between $t = 0s$ to $t = 6.25s$, FIG. \ref{fig:osco_tcl_1}, depicts the condition where the LEDs emulate random behaviour of the TCLs where no control is enforced.

\begin{figure}[t!]
\centering
\includegraphics[scale=0.3]{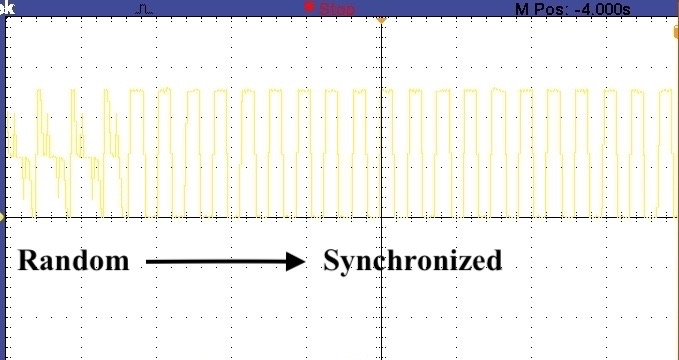}\\
\caption{Transition of LEDs from the random period to synchronized state.}
\label{fig:osco_tcl_2}
\end{figure}

\begin{figure}[t!]
\centering
\includegraphics[scale=0.3]{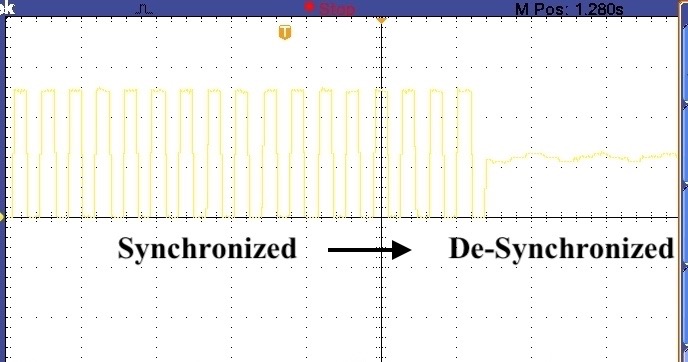}\\
\caption{Transition of LEDs from the synchronized period to desynchronized state.}
\label{fig:osco_tcl_3}
\end{figure}

\par After $t = 6.25s$, the phases of the LEDs get synchronized. Following \eqref{eq10}, LEDs oscillate with their mean frequency. During this period, the LEDs have no phase difference between them. FIG. \ref{fig:osco_tcl_2} shows the synchronized state of the LEDs. The synchronized state of the TCLs is a highly undesirable condition in a power system.

\par Further, succeeding $t = 12s$, the LEDs induce phase difference of $\alpha = \pi/2$ between them. As seen from FIG. \ref{fig:osco_tcl_3}, the output of the LEDs are minimized. Additionally, the fluctuations are also minimized.

\par The results of the hardware implementation were in conjunction with the results of computer simulations. Thus, the results of the proposed theory are validated.

\section{Competitive Benchmark}\label{sec 6}

\par In this section, the authors compare the results obtained from the proposed model with the models mentioned in literature. From FIG. \ref{fig:power_hetero_tcl_1}, the $P_{agg}$ is $750$kW ($45.18\%$ of maximum power consumption). The fluctuations are quantified using root mean square error ($RMSE\%$) given as follows,

\begin{equation}
\begin{aligned}
%RMSE &= \sqrt{\frac{1}{T}\frac{$\int_{0}{T}($P_{ref}-$P_{agg}$$)$dt}{P}}
% RMSE = \sqrt{\frac{1}{T}\frac{$\int_{0}{T}($P_{ref}-$P_{agg}$$)dt${$P_{base}^2$}
RMSE \% &= \sqrt{\frac{1}{T} \frac{\int_{0}^{T}\left(P_{ref}(t) - P_{agg}(t)\right)^2dt}{P_{base}^2}} \times 100.
\end{aligned}
\label{eq12}
\end{equation}

\par For a heterogeneous ensemble of TCLs, the relative error computed is within $2.49\%$, and the $RMSE\%$ is $4.8685\%$. Juxtaposing these results alongside with the minimum variance law (MVC) \cite{callaway2009tapping}, the latter achieves relative error lesser than $5\%$.
\par TABLE \ref{tb:methods} shows the efficacy of a distributed averaging model against several state-of-the-art methods.
The relative error obtained in reference to PRC based technique \cite{bomela2018} were $4\%$ to $8\%$.  Whereas compared to the statistical model presented in the literature\cite{koch2011modeling}, the $RMSE\%$  of the presented model is marginally higher. As compared to the Boolean phase oscillator model \cite{bajaria2019dynamic}, the relative errors are similar. Thus, the distributed averaging model achieves competent results while avoiding computational complexities.

\begin{table}[h!]
\begin{center}
\caption{Comparison with state-of-the-art methods}\label{tb:methods}
\begin{tabular}{ ccc} 
Parameter & Relative error$\left(\%\right)$ & RMSE $\left(\%\right)$\\\hline
\hspace{1cm}\cite{callaway2009tapping} & $5\%$ & $1\%$ \\
\hspace{1cm}\cite{vrettos2012} & $30-50\%$ & $1.18-8\%$\\ 
\hspace{1cm}\cite{bomela2018}& $4-8\%$ & $2.31-5.89\%$\\
\hspace{1cm}\cite{koch2011modeling}& - & $2.27\%$\\
\hspace{1cm}\cite{bajaria2019dynamic}& $2\%$ & $6.3\%$\\
Distributed Averaging & $2.49\%$ & $4.8685\%$\\\hline
\end{tabular}
\end{center}
\end{table}

\section{Data-driven Approach to Phase Desynchronization of TCLs}

\par As discussed in our previous work \cite{bajaria2019dynamic}, delay calculator (DC) is the underlying theory behind choosing the right phase value so as to achieve controlled desynchronization. In this section, we further exploit the benefits of DC by analysing it for higher dimensions. It can be noted though, DC provides accurate results to deploy load following (refer section \ref{sec 5}) with a sectional phase distribution. But, what if DC needs to be evaluated for all the possible permutations and combinations? It is quite obvious that, as the number of dimensions increase, the volume of data blow up exponentially making it difficult for analysis, commonly referred to as ``curse of dimensionality (COD)" \cite{bellman1966dynamic}. In order to avoid COD and provide a different perspective to DC for controlled desynchronization of TCLs, we propose a newer machine learning (ML) based model for the same. Out of the available sets of ML techniques, we try implementing DNN for the discussed case. An obvious choice is to use a multi-variate regressor, but to our analysis DNN proves to be more accurate. Also, DNN must be a right choice taking into consideration COD and making use of the available model/data for a simple application like TCL. There is an immense amount of literature available on DNNs, in order to understand the concepts and terminologies of DNNs (out of scope of this context, but the reader can refer the citation) \cite{goodfellow2016deep}. 

\begin{figure}[t!]
\centering
\begin{tabular}{c}
\includegraphics[scale=0.17]{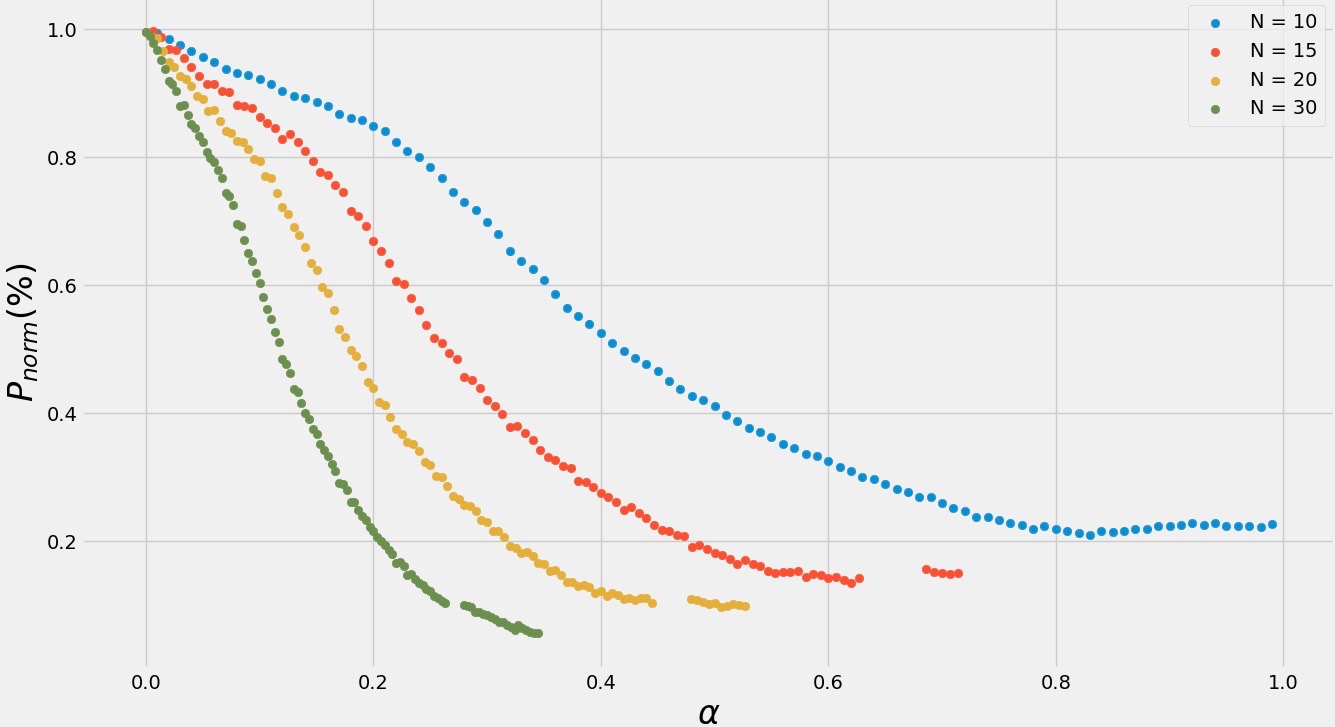}
\end{tabular}
\caption 
{Variation of $P_{norm}$ vs $\alpha$ across $N$.}
\label{fig:Palpha}
\end{figure}

\subsection{Generating Dataset}
\par In order to train the supervised learning model, for predicting the value of $\alpha$, we generate a dataset using parameters mentioned in TABLE \ref{tb:tab2}. For emulating a practical environment, we choose the value of $P$ and dutycycle from a random distribution. The number of TCLs vary from $N = 10$ to $N = 500$. Additionally as the value of $P_{agg}$ varies from a maximum at $\alpha = 0$ radians, to a minimum at $\alpha$ = $2\pi/N$ radians, we compute $P_{norm}$ using equation \ref{eq12} for $100$ equally spaced values of $\alpha$ between $0$ and $2\pi/N$ radians. As evident from FIG. \ref{fig:Palpha}, the value of $P_{norm}$ varies between maximum and minimum.
\par This dataset encompasses the parameters of most commercial and household TCLs and gives us a sufficiently large data to train our DNN for predicting values of $\alpha$ for any $N$. The dataset contains $49490$ data points with two independent variables $N$ and $P$, and one target variable $\alpha$. 

\subsection{Data Preprocessing}
\par The dataset generated using feature engineering generally do not have missing value or $NaN$ and so goes for the data generated using DC theory. However, we need to normalize the dataset prior to model building and training for improved accuracy.
\par For normalization, we use minmax normalization to linearly transform the dataset. Minmax transformation transforms each data point based on the minimum and maximum value along a particular feature. It can be expressed as,

\begin{equation}
\begin{aligned}
 X_{norm} = \frac{(X - X_{min})}{(X_{max} - X_{min})}
 \label{minmax}
\end{aligned}
\end{equation}

Where $X_{norm}$, $X_{min}$, and $X_{max}$ are the normalized value of the data point $X$, minimum and maximum value across the feature $X$, respectively. To validate our model we segregate the dataset into training and testing data in the ratio $7:3$.

\subsection{Model Training and Results}
\par Next, we train the DNN using dataset generated in previous sections. In order to optimize computational cost and standardization of code, we make use of scikit-learn \cite{pedregosa2011scikit} and keras \cite{chollet2015keras} libraries in Python $3.7.3$. A simple $4$-layered DNN with $2$ dense hidden layers comprising of $2$ and $4$ nodes, respectively has been implemented. With an appropriate choice of activation functions, epochs and batch sizes a RMSE of $3\%$ is attained. The model evaluation metrics are mentioned in TABLE \ref{tb:tab4} showing competitive results and accuracies.

\begin{table}[h!]
\begin{center}
\caption{Model Evaluation}\label{tb:metrics}
\begin{tabular}{ cc} 
Metric &  Value\\\hline
Root Mean Squared Error & $3.06\%$ \\
Mean Squared Error & $0.09\%$ \\
Mean Absolute Error & $ 0.0214^\circ$ ($ 3.7{ e^{-4}}$ radians) \\
\hline
\label{tb:tab4}
\end{tabular}
\end{center}
\end{table}

\begin{figure}[t!]
\centering
\begin{tabular}{c}
\includegraphics[scale=0.17]{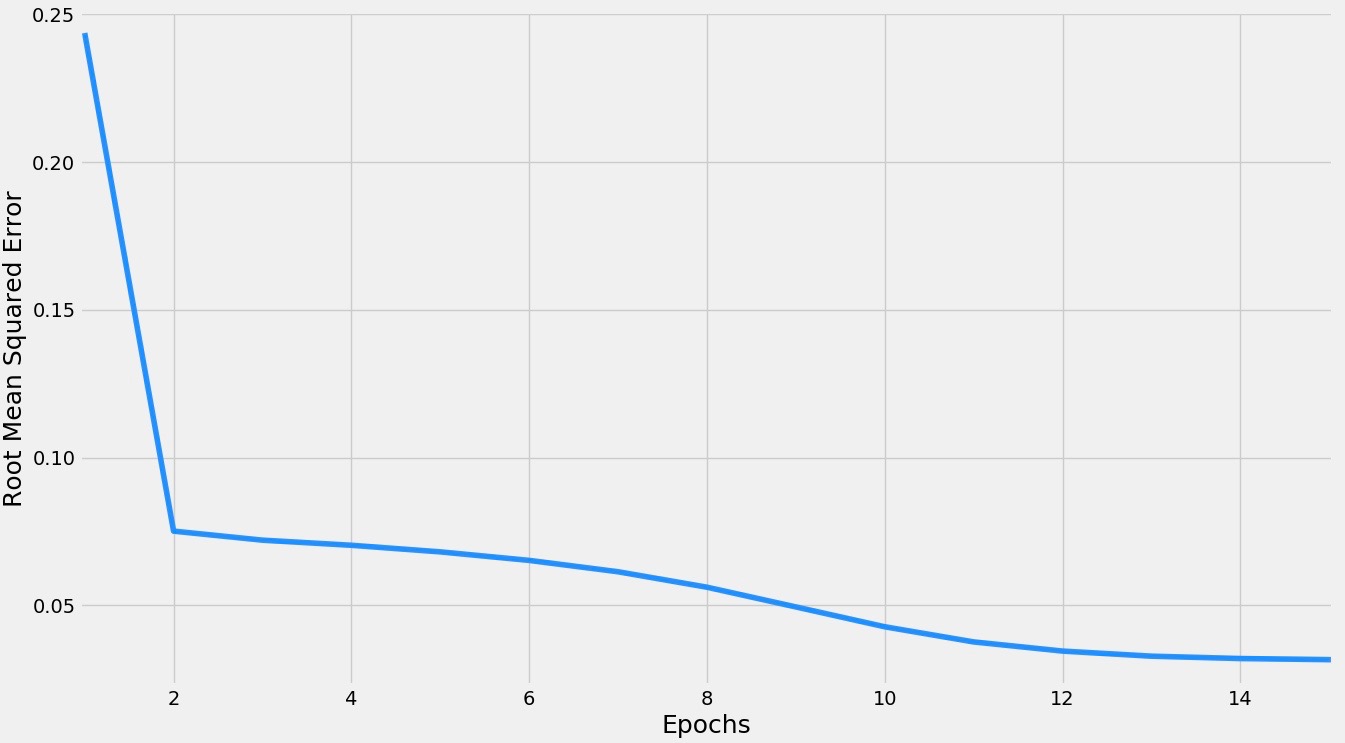}
\end{tabular}
\caption 
{Root Mean Squared Error(RMSE) variation with iterations.}
\label{fig:RMSEvsEp}
\end{figure}

\begin{figure}[t!]
\centering
\begin{tabular}{c}
\includegraphics[scale=0.17]{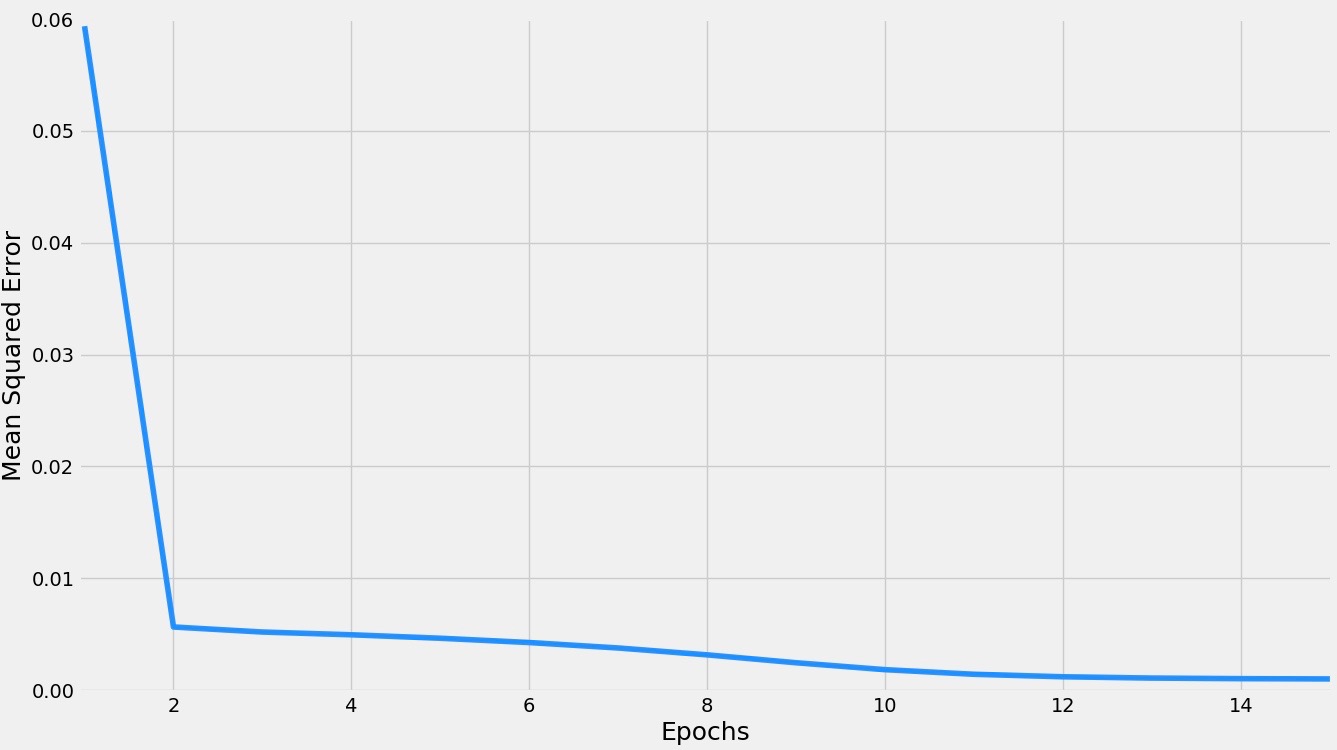}
\end{tabular}
\caption 
{Mean Squared Error(MSE) variation with iterations.}
\label{fig:MSEvsEp}
\end{figure}

\begin{figure}[t!]
\centering
\begin{tabular}{c}
\includegraphics[scale=0.17]{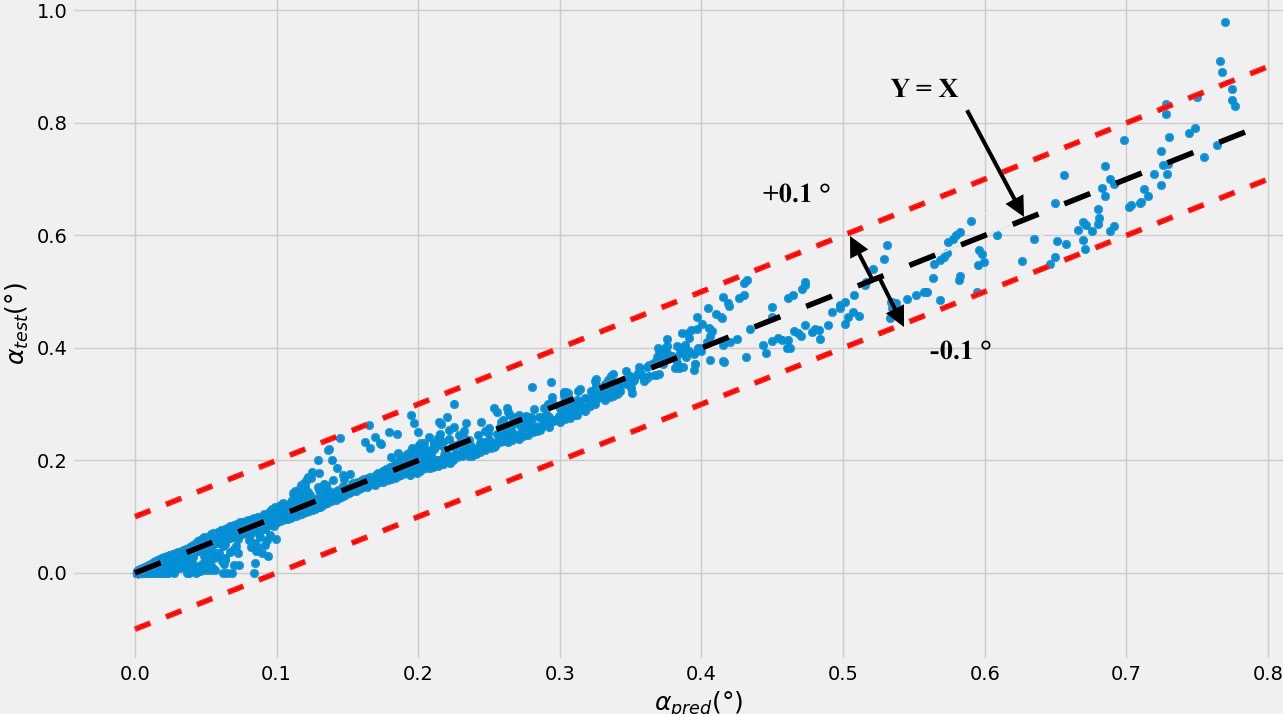}
\end{tabular}
\caption 
{$\alpha_{true}$ vs $\alpha_{pred}$.}
\label{fig:AlphaVSAlphaPred}
\end{figure}

\begin{figure}[t!]
\centering
\begin{tabular}{c}
\includegraphics[scale=0.17]{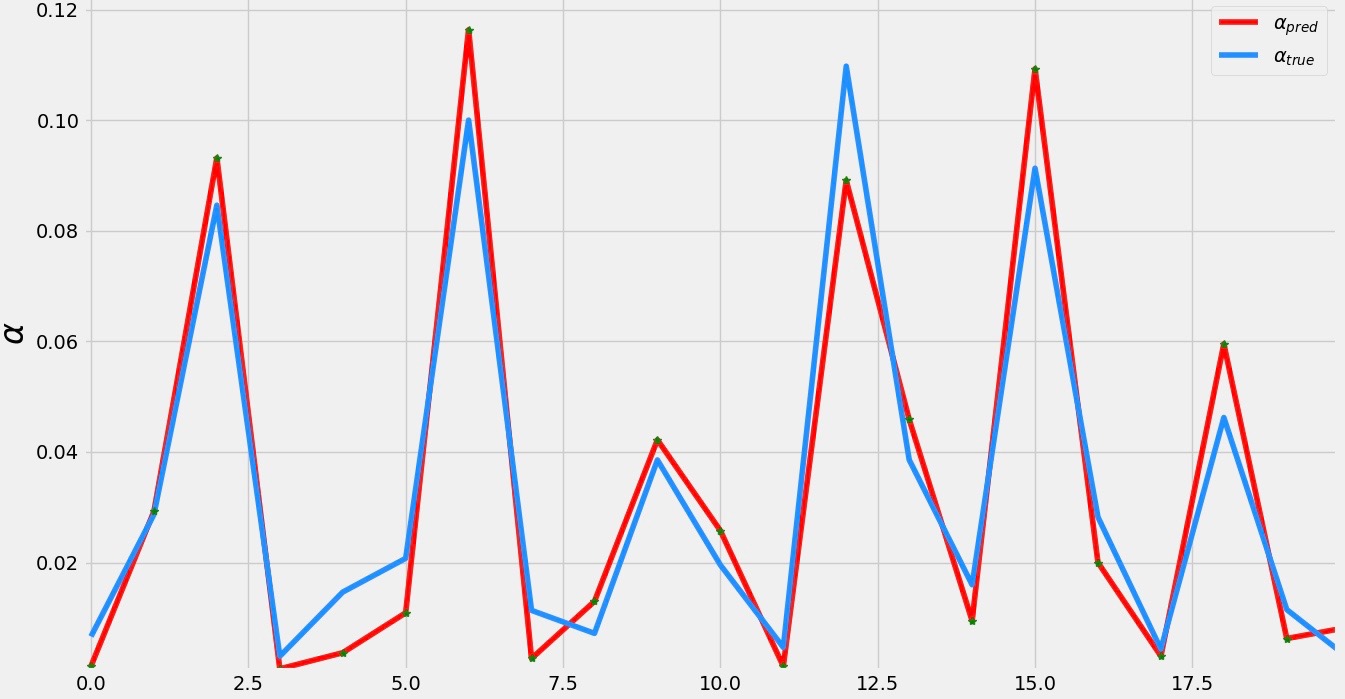}
\end{tabular}
\caption 
{Plot of $\alpha_{pred}$ compared to $\alpha_{true}$.}
\label{fig:Timeseries}
\end{figure}

\par The variation of RMSE, MSE with epochs are shown in FIG. \ref{fig:RMSEvsEp}, FIG. \ref{fig:MSEvsEp}; depicting DNN getting trained along the iterations. It can be observed nearly after $~15$ iterations, the accuracy of the DNN starts to saturate. 

\par We compare the values of $\alpha_{pred}$, $\alpha_{true}$ using a correlation map as shown in FIG. \ref{fig:AlphaVSAlphaPred}. Additionally, as seen from FIG. \ref{fig:Timeseries} and TABLE \ref{tb:tab4} the value of $\alpha_{pred}$ is within $0.1^\circ$($1.7e^{-3}$ radians) of the true value showing the efficiency of predictions.

\section{Conclusion}\label{sec7}
\par The major highlight of this work is the application of a distributed averaging protocol for the desynchronization of a population of TCLs. The system of TCLs are visualized as oscillators, and their power fluctuations have been minimized. A significant cornerstone of the proposed work is that user comfort is unimpaired. The parameters governing the dynamics of the system are then articulated as a set of differential equations.

\par Apart from power fluctuations, aggregation, and user comfort, another breakthrough is the simplicity of the presented model and reduced computation. Load following helps the utility by improving the usability of the generated power, thereby supporting in load-frequency control and maintaining voltage profile. A reduction in power fluctuation and aggregation would ease off the demand pressure from the industries directly or indirectly associated with power generation, distribution, protection, and maintenance. The results of both simulations, as well as hardware, support the findings of the presented works. The DNN model provides a convenient way to compute the value of the delay further saving us computational cost and optimizing the system. The results obtained are motivating and show great potential to be implemented as a plug-n-play device and be made available in the market.

\doclicenseImage

\end{document}